\documentstyle{article}
\parskip=\normalbaselineskip
\def\sane#1{\ifvmode{\leavevmode\hbox{#1}}\else{\hbox{#1}}\fi}
\def\sco{\sane{Sco~X-1}}
\def\about{$\sim$}

\newcommand{\mdot}{$\dot{M}$}
\newcommand{\Sz}{\mbox{S$_{\rm Z}$}}
\newcommand{\Dz}{\mbox{D$_{\rm Z}$}}
\newcommand{\Vz}{\mbox{V$_{\rm Z}$}}
\newcommand{\Az}{\mbox{A$_{\rm Z}$}}
\newcommand{\etal}{{\rm {\it et~al.}}}

\newcommand{\rf}{\par\noindent\hangindent 15pt{}}

\begin{document}

{\Large\bf The Timing Properties of \sco\ along its Z track with EXOSAT.}\\

{\large Stefan W. Dieters$^{1}$ and Michiel van der Klis$^{2}$}\\
$^{1}${\it CSPAR, University of Alabama In Huntsville, Huntsville,
AL 35899, U.S.A.}\\
$^{2}${\it Astronomical Institute, ''Anton Pannekoek'', University of
Amsterdam, Kruislaan 403, 1098 SJ Amsterdam, The Netherlands}\\

{\large\bf Abstract.}

We present a systematic, homogeneous analysis of all the EXOSAT ME, high time
resolution data on \sco.  We investigate, for the first time, all power
spectral properties of the $<$100\,Hz quasi-periodic oscillations (QPO) and noise of
\sco\ as a function of position on the Z-shaped track traced out in the X-ray
colour-colour diagram. For this purpose, we introduce a new generally
applicable method for parametrizing the position of a source on its track in a
colour-colour or hardness-intensity diagram. Generally, the properties of
\sco\ vary smoothly as a function of position along the Z track. However, some
variability parameters change abruptly at either of the vertices of the Z
track indicating that the branches of the Z track represent distinct source
states not only in spectral state but also in rapid variability
characteristics. All variability components are found to have energy spectra
harder than the average flux. We show that the very low frequency noise (VLFN)
is consistent with being solely due to motion along the Z track. The power
spectra of the X-ray intensity as well as source position along the Z track
extend, unbroken, to time scales of nearly 1~day. We study the high frequency
noise (HFN) component for the first time in sufficient detail to show that
there are changes in the HFN cut-off frequency with position on the Z track.
It changes abruptly from $\sim$\,75\,Hz to $\sim$\,35\,Hz at the
normal/flaring branch vertex. The HFN is found to extend out to about 300\,Hz.
The QPO show a remarkably rapid change in frequency at or just before the
normal-branch/flaring-branch (NB/FB) vertex. This transition happens within
1.5\% of the entire extent of the Z track. The QPO themselves are visible for
17\% of the Z. We find a new sort of behaviour near the NB/FB vertex, i.e.,
rapid excursions from the NB into the FB and back again taking only a few
minutes. We found several indications that position on the Z track is not the
only parameter governing the behaviour of \sco. The most dramatic examples of
this are two brief episodes where the QPO frequency changed rapidly without
the usual changes in colours and intensity which accompany a change of source
state. In one case the frequency rapidly rose from 6\,Hz to 16\,Hz, and then
returned to 6\,Hz, while \sco\ apparently remained unmoved on the normal
branch in the colour-colour diagram. In the second case the QPO frequency
changed from 16\,Hz to 7\,Hz without the usual indications of a passage
through the normal/flaring branch vertex (simultaneous dips in the count rate
and colours as the frequency changes through 8\,Hz). Thus it seems that
deviations from the usual one to one correspondence between QPO behaviour and
spectral state, perhaps due to QPO frequency mode switching do occasionally
occur.

---stars:\,individual:\,\sco---stars:\,neutron:\,Z-sources---X-rays:\,stars

{\large\bf Thesaurus Codes.}

02.01.2---03.09.1---08.02.1---08.09.2:\,\sco---08.14.1:Z-sources---13.25.5

{\large\bf 1: Introduction.}

\sco\ is a low mass X-ray binary (LMXB). These systems consist of a neutron
star or a black hole and an accretion disk fed by Roche lobe overflow from a
low mass ($\leq$2 M$_{\odot}$) companion. The optical and X-ray fluxes of
LMXBs are dominated by processes in or near the accretion disk. Persistently
bright LMXBs such as \sco\ are thought to contain a neutron star, because  two
similar systems show type I bursts (i.e., GX~17+2; Sztajno \etal\ 1986,
Kuulkers \etal\ 1997 and Cyg~X-2; Kuulkers \etal\ 1995). In the cases of \sco\
and Cyg~X-2 the companions must be evolved, as the orbital periods are long
(19.2~hr and 252~hr, respectively) and the inferred mass accretion rates are
high. No X-ray pulsations have been detected from any of the persistently
bright LMXBs (Vaughan \etal\ 1994). 

Since its discovery (Giacconi \etal\ 1962) \sco\ has been extensively observed
in many different wavelength bands (e.g., correlated optical/X-ray photometry:
Mook \etal\ 1975, Petro \etal\ 1981, Augusteijn \etal\ 1992, Mc\,Namara \etal\
1994, optical spectroscopy: La\,Sala and Thorstensen 1985, radio: Hjellming
\etal\ 1990, UV: Vrtilek \etal\ 1990, 1991, hard X-rays: Jain \etal\ 1984,
Ubertini \etal\ 1992). The discovery (van der Klis \etal\ 1985, Hasinger
\etal\ 1986, Middleditch \& Priedhorsky 1986) of rapid (\about15--55\,Hz) and
slow ($\sim$6\,Hz) quasi periodic oscillations (QPO) in the X-ray flux of the
brightest LMXBs opened a way to a greater understanding of the accretion flow
close to the compact object in these systems. Observations show that the
properties of the QPO and associated noise components are related to the
source's spectral state. The spectral state is most easily determined by using
an X-ray colour-colour diagram (CD). On such a diagram six of the brightest
LMXBs trace out a distorted Z shaped path; they are therefore called 'Z'
sources. These sources are GX~5$-$1, Cyg~X-2, GX~340+0, GX~349+2, GX~17+2, and
\sco. 

The three branches of the Z are called the horizontal branch (HB: topmost
stroke), the normal branch (NB: the downward stroke) and the flaring branch
(FB: the bottom stroke). The branch names are mostly historical; the
horizontal branch is not necessarily horizontal, the source may or may not
spend most of its time in the normal branch, and only \sco, GX~17+2 (Penninx
\etal\ 1990) and GX~349+2 (Ponman \etal\ 1988) show flares in the flaring
branch.  The normal and flaring branches of \sco\ correspond to the well
documented (e.g., Mook \etal\ 1975, Petro \etal\ 1981 and Hertz \etal\ 1992)
quiescent and active states seen at both optical and X-ray wavelengths
(Augusteijn \etal\ 1992, Hasinger 1987a, Priedhorsky \etal\ 1986). 

There are six distinguishable components in the power spectra of Z sources
(e.g.\/ van der Klis 1989a, 1995a,b). The very low frequency noise (VLFN) is
well fitted by a power law. It dominates the power spectrum below 0.1\,Hz.
High frequency noise (HFN) is approximately flat out to a frequency near
50-100\,Hz and then slopes down towards higher frequencies. It has been fitted
with either an exponentially cut-off power law or a Lorentzian peak centered
at zero frequency. The low frequency noise (LFN) either appears as a component
that is decreasing monotonically with frequency (``red noise") and extends out
to about 10\,Hz; or in other sources it has the shape of a very broad peak
with a central frequency near 2\,Hz. It has been represented by an
exponentially cut-off power law (red or peaked depending on the sign of the
power law index), or by a Lorentzian peak centered at zero frequency (red), or
near 2\,Hz (peaked). Finally, there are three distinct types of QPO; one
associated with the horizontal branch with frequencies between 5 and 55\,Hz,
another associated with the normal and flaring branches (6-20\,Hz) and a pair
of kHz frequency QPO (see van der Klis 1998 for a review of kHz QPO). This
paperdeals with the first two metioned types of QPO only.   
These QPO have been fitted by Lorentzian or Gaussian peaks. 

Both the VLFN and the HFN are observed in all branches. The LFN is observed
only in the horizontal branch, and in the very upper part of the normal
branch. Also observed in this part of the Z track are intensity dependent
\about15--55\,Hz QPO, the horizontal branch QPO (HBO). When HBO are observed,
LFN is always present as well. On the mid- to lower NB another distinct type
of QPO is observed. Occasionally these normal branch oscillations (NBO) occur
simultaneously with the HBO. The NBO frequency remains approximately constant
at a frequency near 6\,Hz despite 30\% changes in intensity when the source
moves up and down the normal branch. In \sco\ (Priedhorsky \etal\ 1986) it has
been observed that when the source moves into the lower part of the FB, the
QPO frequency increases from 6\,Hz to at most $\sim$\,21\,Hz. Further up the
FB the QPO melt into a continuum of HFN. These high frequency QPO are called
flaring branch oscillations (FBO). They have also been seen in GX~17+2
(Penninx \etal\ 1990, Kuulkers \etal\ 1997). The FBO are thought to be closely
related to the NBO because of the continuous transition from one to the other
(the QPO found in the flaring branch of Cyg~X-2, Kuulkers \& van der Klis
1995, may be another phenomenon). The properties of the various power spectral
components are summarized in Table\,1 of van der Klis (1989a) and Fig.\,1 of
Lamb (1989). See also the reviews of Lewin, van Paradijs, and van der Klis
(1988) and van der Klis (1989a,b, 1995a,b, 1998). 

The tight correlation of all observables with the instantaneous position of
the source on the Z-track and the continuous nature of the motion along the Z
track suggest that there is only one underlying physical parameter
responsible. The obvious candidate is the mass accretion rate, $\dot{M}$. In
\sco\ and Cyg~X-2 it is observed that the ultraviolet-line and -continuum
fluxes (Vrtilek \etal\ 1990,1991, Hasinger \etal\ 1990) and the optical
brightness (Augusteijn \etal\ 1992) all increase as the source moves from the
left-most end of the horizontal branch along the Z track toward and up the
flaring branch. This indicates that $\dot{M}$ increases along the Z track in
this sense (Hasinger \etal\ 1990). The NB/FB transition is thought to occur
near the Eddington limit (Lamb 1989). Note that the X-ray count rate is not a
good measure of $\dot{M}$ as it increases in both directions away from the
NB/FB vertex, up the NB as well as up the FB. 

Different models have been proposed for the HBO and the NBO/FBO. The most
successful explanation for the HBO is given by the ``beat-frequency model" of
Alpar and Shaham (1985; also Lamb \etal\ 1985). Here matter is
quasi-periodically removed by the neutron stars' magnetic field from the inner
edge of the accretion disk. According to the model of Fortner \etal\ (1989)
for the the NBO/FBO, these oscillations originate in a cool radial inflow
surrounding the neutron star and its magnetosphere. This region is prominent
only at luminosities near the Eddington luminosity ($L_{Edd}$) when, as
already pointed out by Hasinger (1987a), radiation force significantly slows
the in-falling material. Under these circumstances opacity-driven
instabilities can develop with properties consistent with those observed in
the NBO. According to Lamb (1989), near $L_{Edd}$ the flow becomes unstable
and photo-hydrodynamic oscillations are excited with frequencies ranging from
near 6\,Hz to greater than 10\,Hz, as seen in moving from the NB to the FB. A
different model for NBO/FBO, where these oscillations are caused by sound
waves in a thick, torus-like inner accretion disk, has been proposed by Alpar
\etal\ (1992). 

Compared to Cyg~X-2 and GX~5-1, the sources \sco\ and GX~17+2 have short
horizontal branches which are oriented nearly vertically. No HBO were detected
using EXOSAT on the HB of \sco\ (which itself was observed only once;
Hasinger, Priedhorsky and Middleditch 1989, hereafter HPM89). However van der
Klis \etal\ (1997) report, using RXTE observations, a QPO feature with
properties similar to HBO in \sco. The LFN of \sco\ and GX~17+2
(Langmeier \etal\ 1990, Penninx \etal\ 1990, Kuulkers \etal\ 1997) has a peak
near 2\,Hz, whereas in Cyg~X-2, GX~5$-$1 and GX~340+0 it has a ``red"
(monotonically decreasing with frequency)  shape. \sco, GX~17+2 and GX~349+2
have prominent flaring branches with classic flaring behaviour, while GX~5-1
and Cyg~X-2 have small FBs. In these two  sources, the count rate dips, rather
than flaring, when the source moves into the FB. GX~340+0 also has a weak FB,
with a combination of both dipping and flaring behaviour (Penninx \etal\ 1991,
Kuulkers \& van der Klis 1996). While FBO have been observed from \sco, and
GX~17+2, they are weak for GX~340+0 (Penninx \etal\ 1991, Kuulkers \& van der
Klis, 1996), or absent on the flaring branches of GX~5-1 (Kuulkers \etal\
1994). These systematic differences suggest that there are two types of Z
source, or that \sco\ and Cyg~X-2 are near the two extremes of a continuous
range of behaviour differences, possibly governed by inclination angle
(Hasinger \&\ van der Klis 1989, Kuulkers \etal\ 1994, 1997), or neutron star
magnetic field strength (Psaltis, Lamb \& Miller, 1995). 

Most of the observational material gathered with EXOSAT on \sco\ has been
previously analysed (Middleditch \& Priedhorsky 1986, Priedhorsky \etal\ 1986,
Pollock \etal\ 1986, van der Klis \etal\ 1987, HPM89). Ginga data on \sco\
were reported by Hertz \etal\ (1992). These studies concentrated upon
individual observations and hence were necessarily fragmentary and without the
benefit of our current more global view of Z source behaviour. Here, 
we present the first comprehensive and homogeneous
analysis of the EXOSAT \sco\ data. As part of a project to look for rapid
power fluctuations, we examine these data in much greater detail than in
previous studies. Particular attention is paid to freezing any changes in QPO
properties associated with changes in brightness or position upon the
colour-colour diagram. We introduce a new method for parametrizing the
position along the Z track, which can be applied to colour-colour tracks of
any shape. We study the variation of all power spectral components as a
function of position along the Z track for all branches in a consistent
manner. The behaviour of the QPO near the NB/FB vertex is closely examined,
and three different types of behaviour are identified. A previously
unrecognized, rapid form of the NB/FB transition is noted. Near the NB/FB
transition, the QPO frequency is found to be very sensitive to position along
the Z track: the change from 7.5\,Hz to 12\,Hz occurs within 1.5\% of the span
of the Z track. We present two cases where \sco\ seemed not to follow usual
Z-source behaviour of smooth motion along the Z track with strictly correlated
changes in fast variability properties. 

{\large\bf 2: Observations and Analysis.} 

\sco\ was observed several times with the EXOSAT satellite during the period
1983 to 1986. We used data taken with the medium energy (ME) instrument
(Turner, Smith and Zimmermann 1981, White and Peacock 1988). The array of 8 ME
detectors was arranged in two half arrays of 4 detectors, i.e., half 1:
detectors 1--4 and half 2: detectors 5--8. Each detector consisted on an argon
filled proportional counter atop a xenon filled proportional counter separated
by a thin mylar window. The Ar counters were sensitive over the 1 to 20\,keV
range while the Xe counters were sensitive over the 5 to 50\,keV range. There
were 4 thick (numbers 1,2,5,6) and 4 thin (3,4,7,8) windowed Ar detectors;
these had substantially different energy responses. Each counter could be
individually switched on or off. Each half array of detectors could be pointed
in slightly different directions. The On Board Computer (OBC) could be
programmed to collect, simultaneously, data in several modes. These modes were
separated into High Time Resolution (HTR) and High Energy Resolution (HER)
modes. We examined all observations, but we concentrated on observations where
energy resolved data and high time resolution data were simultaneously
available. An observation log is shown in Table\,1. The background in the Xe
counters was always near 75\,cts/sec/counter while the Ar background was much
lower, about 12\,cts/sec/counter. We did not use the early 1983 observations
because of the many detector set-up changes and the low time resolutions used
in both the HER and HTR modes. 

The most common set-up used for observing \sco\ was with one half array with
only the argon counters switched on, and the other half array with only the
xenon counters switched on. The Xe half (array half 2) was in all observations
pointed directly at \sco\ while the Ar half (array half 1) was usually offset
to a collimator response of 8--12\%. This was done to avoid damage to the Ar
counters from high count rates. Generally the argon counters were used to
collect data in 4 energy channels at time resolutions between 4 and 16\,ms
(HER7 mode, I7 data) while the xenon counters were used to produce data in one
energy channel with a 1 or 2\,ms time resolution (HTR3 or HTR5, T3 or T5
data).  The setup was different on Feb 25, 1985, when data were collected from
all eight Xe counters, which were pointed directly at \sco, but no Ar data
were collected. Another exceptional observation (3 Aug 1984) had all 8 (Ar+Xe)
detectors on and used a higher collimator response than usual (12--23\% rather
than 8--12\%). During this observation the OBC was programmed to give the
combined Ar+Xe detector count rate every 8~ms (HTR3, T3 data) and Ar and Xe
multichannel energy data every 32\,ms (HER5 mode, E5 data).  The resulting
count rates were $\sim$5900\,cts/sec in the NB, reaching a maximum of
9800\,cts/sec in the FB. The peak count rate is the highest recorded in any
EXOSAT observation.
 
The energy boundaries of the four I7 channels normally used were 0.9, 3.1,
4.9, 6.6 and 19.5~keV. The four Ar bands defined by these boundaries we will
call bands 1 to 4, respectively. We converted the Aug 1984 E5 data into 4
channel data with the same energy boundaries as the I7 data.

When we report Xe count rates the numbers will always refer to background
corrected count rates over the full Xe energy range and reduced to counts per
whole array (8 detectors) at 100\% collimator response. The deadtime
corrections for the Xe HTR mode data were small ($<$1\%) and here will provide
a valuable framework in which to assess more recent RXTE observations. 
The Ar count rates are background and deadtime corrected (Andrews
\& Stella 1985) and scaled to a 10\% collimator response for 8 detectors.
Although the latest collimator response calibration (Kuulkers \etal\ 1994,
Kuulkers 1995) was used, the collimator responses at the large offsets used
for \sco\ with the Ar detectors are somewhat uncertain. The systematic 
uncertainty in the
corrected Ar count rate is $\sim$5\% varying from one spececraft pointing to
the next.

We define the ``soft colour" as the ratio of the background subtracted and
deadtime corrected count rate in band 2 to that in band 1; the ``hard colour"
is the ratio of band 4 to band 3. This choice of colours gave the most
separated normal and flaring branches in the colour-colour diagram. 

There are systematic uncertainties in determining the colours. Uncertainties
in determining the background result in changes of $<$0.14\% in the soft
colour and $<$4.7\% in the hard colour.  The colours as measured for the Crab
are lower using the thin windowed detectors than using the thick windowed ones
(Kuulkers \etal\ 1994, Kuulkers 1995).  On 20 Aug (day 232) 1985, just a few
days before the Aug 1985 observations, detector 3 of array half 1 failed
(Haberl 1992). This was a thin windowed detector and so there is a marked
increase in measured soft colour after the failure.  The increase was 2.5\%
for the Crab and $\sim$1.5\% for GX~5-1. There are also changes of about 1\%
in both the soft and hard colours due to small changes in relative detector
gains. 

{\it 2.1: Determining the Z-track position: \Sz.} 

The tight correlation of all source properties with position along the
Z-track, and so by inference with $\dot{M}$, suggests that it is reasonable to
measure all observables as a function of Z-track position. This was first done
using a ``rank number'' by Hasinger \etal\ (1990). Separate observations were
ranked according to position. Because the HB/NB (hard) and the NB/FB (soft)
vertices marked state changes these where assigned special rank numbers. The
measurement of Z-track position was refined by Hertz \etal\ (1992) who fitted
a tilted  paraboloid and measured the Z-track position as an arc away from the 
NB/FB vertex. Here, we extend the measurement of position along the Z to
arbitrarily shaped tracks. We develop an automatic and reliable method to
separate data on closely spaced branches. We introduce a normalization which
makes it easier to compare the Z tracks of different sources. 

The position within the Z of each point in the colour-colour diagram was
parametrized by drawing a smooth curve through the colour-colour points, and
projecting the points onto this curve. The smooth curve was defined by
choosing, by hand,  normal points in the CD. These normal points were ranked.
Two cubic spline fits were then made, one for each colour, between rank and
colour. The two resulting splines form a numerical representation of the Z
track in the colour-colour plane. 

For each colour-colour point, $P$, an initial position on the Z track was
defined as the point, $Q$, on the track closest to $P$. The arc length along
the Z track between $Q$ and the  normal/flaring branch vertex was calculated.
The NB/FB vertex was chosen as a reference point because it is the most
clearly defined: it is the point on the Z track with the lowest soft and hard
colours. The arc length between Q and the NB/FB vertex depends upon the scale
of the CD, which in turn depends upon the choice of energy bands. The
parameter \Sz\ was defined as the arc length along the Z normalized such that
the HB/NB vertex corresponds to \Sz=1 and the NB/FB vertex to \Sz=2. Our
definition of \Sz\ is similar to that of ``rank number" of Hasinger \etal\
(1990) in that the values 1 and 2 are assigned to the vertices, and to that of
``s$_z$" of Hertz \etal\ (1992) in that we use arc length along the Z track to
determine position. Following Hertz \etal\ (1992) we call the distance between
$P$ and $Q$, normalized in the same fashion as \Sz, \Dz. So, the above
procedure basically transforms the two colour coordinates (soft and hard
colour) into the two coordinates \Sz\ (distance along the Z track) and \Dz\
(distance perpendicular to the Z track).

When comparing Z-track positions from different detector and different
sources there are unavoidable dependencies upon the choice of energy bands
used, the detector response and the source's energy spectrum. With our
normalization  these dependencies are much reduced; the major factor remaining
is  the relative ranges of the colours. In our case the ranges are nearly
equal.

In determining \Sz\ there is a complication. As can be seen from Fig.\,1, the
normal and flaring branches are relatively close together. The statistical
uncertainties are sufficient that some colour-colour points which belong on
one branch, are actually closer to the other branch. At the 196-sec
integration time that we used in Fig.\,1 such wayward \Sz\ measurements are
rare and therefore easily identified. However, \Sz\ shows changes on
time scales less than few hundred seconds which are smeared with 196-sec
integrations. In an attempt to follow these faster changes in \Sz, we also
used shorter (16 and 32\,sec) integration times. Then, due to the larger
statistical errors, the procedure outlined above created two sets of \Sz\
values, the majority corresponding to the branch on which \sco\ actually was,
and the other set incorrectly attributed to the adjacent branch. Because of
the rough symmetry about the vertex, the two sets of \Sz\ values approximately
mirrored each other about the line \Sz=2 when plotted against time. We tested
several automatic methods to identify the wrongly assigned points and move them
back to the correct branch.

We first tried to identify wayward \Sz\ values by finding values that  were
more than a set number of standard deviations from the mean of their time-wise
neighbours. The local standard deviation and mean of \Sz\ were calculated from
the nearest 3--5 neighbours on each side of the value being tested. Using this
method, it was possible to distinguish isolated wayward points, but the method
totally failed when there were 3 or 4 wayward points close together in time.
We then experimented with a method where each 16 or 32-sec \Sz\ value was
compared with a linear interpolation between the 196-sec values. This method
was more effective as it was very good at identifying mis-assigned points, but
it often smoothed out short ($\sim$600\,sec) but real changes in position on
the Z track. The method we finally settled upon, used the {\it mode} of groups
of usually 12 consecutive 16 or 32-sec \Sz\ values   to determine the
long-term trend. Each \Sz\ value that was  more than 3$\sigma$ from the
long-term trend was identified as wayward. The standard deviation $\sigma$ was
determined from the scatter of all 16 or 32 sec points around the long term
trend. 

For each colour-colour point, P, there are up to three possible positions upon
the Z track, corresponding to three local minima in the distance between the
point and the track. As described previously, initially the \Sz\ value
corresponding to Q, the closest position on the Z track (corresponding to the
global minimum) was used. If a point was identified as wayward, then of the
three possible \Sz\ values the one closest to the long term trend was selected
to replace the initial value. This checking and replacing procedure could be
done recursively. Generally only one iteration was necessary. 

The scaling of the \Sz\ values was provided by measuring the arc length
between the HB/NB and the NB/FB vertices. The position of the NB/FB vertex
could be accurately  pin-pointed because it corresponds (by definition) to a
minimum in both colours and because of the local symmetry of the Z-track about
the vertex. The position of the HB/NB vertex (\Sz$=1$ in Fig.\,1) is  harder
to determine than that of the NB/FB vertex because it is  more rounded. The
process used to find the initial, uncorrected \Sz\ values led to clustering of
points on the convex side of the vertex, near the inflection point of the
smooth curve. This is a geometric effect that, incidentally, is markedly
reduced by the procedure used to reassign wayward \Sz\ values. This clustering
was used as a further measure of the NB/FB vertex position and was used to
define the exact arc length of the HB/NB vertex. Its position is in the region
expected from the changes in the power spectra. 

{\it 2.2: Power Spectra.} 

All (primarily Xe) data with 31.25 ms or better time resolution were divided
into segments of 16 or 32 sec in length with no data windowing applied. Power
spectra, normalized as per Leahy \etal\ (1983) were calculated from these data
segments using the FFT algorithm. These  segments had 4096, 2048 or 1024 data
points depending on time resolution. These were combined into dynamical power
spectra binned in frequency to a resolution of about 0.25\,Hz. The evolution
of the QPO could be followed in these dynamical power spectra. 

The power spectra were averaged over hand-picked contiguous sections of the
data where both count rate and \Sz\ were constant to within 10\%. Sometimes,
when the count rate was changing too rapidly for this, the data were split into
sections that reflected the nature of the variation. For example, in a flare,
one or more sections covered the rise, one covered the peak, and one or more
covered the decline. The aim of splitting up the data in this fashion was to
increase sensitivity to any count rate or \Sz\ dependent changes in the QPO. 

When QPO were present in the dynamical power spectrum, the average power
spectra were fitted over the 2--40\,Hz range with a Lorentzian representing
the QPO peak plus a constant level representing the white noise (``Poisson
noise") and any other broad-band components in that frequency range. The power
spectra of data sections near (in time or count rate) those with obvious QPO
were examined for possible QPO peaks. Any suspected QPO peak was fitted.
However, the resulting fit was rejected if the QPO peak(s) was not
significant. 

We then performed a second power spectral analysis, aimed primarily at
measuring the non-QPO power spectral components. In order to make it possible
to measure the VLFN component, we used 128-sec long data segments in this
analysis. No data windowing was applied. The original full time resolutions
were used so that the extent of the HFN could be ascertained as well as
possible. From the fits to the average power spectra of the 16 or 32-sec data
segments, the \Sz\ range where QPO were evident was already clear. The power
spectra calculated from 128-sec segments from outside this range, so without
QPO, and which had similar \Sz\ values, were averaged. In averaging, each
power spectrum was given a weight proportional to how much it overlapped, in
time, with the chosen range in \Sz. These averaged power spectra were fitted
with a model consisting of a VLFN (power law: $\nu^{-\alpha_{VLFN}}$) and an
HFN (exponentially cut-off power law:
$\nu^{-\alpha_{HFN}}e^{-\nu/\nu_{cut_{HFN}}}$) component. When a fit's reduced
$\chi$$^{2}$ value was high and the residuals indicated a further component, a
fit function containing an extra LFN (cut-off power law) and/or Lorentzian
component was fitted. The white noise level in all these fits was fixed at the
level predicted by the theoretical deadtime effect on the Poisson noise (see
Sect.\,2.3). 

In those 128-sec intervals where QPO were evident, we performed fits with a
full VLFN+ HFN+ QPO model. We found that selecting data by \Sz\ was not
sensitive enough to freeze changes in QPO frequency near the vertex or along
the FB. Using the fits to the power spectra of the 16 or 32-sec data segments
as a guide, power spectra from the 128-sec segments where the QPO frequency
was the same to within 1\,Hz were selected and then averaged. No LFN or extra
QPO component were warranted. 

In all these fits, the functions describing QPO and broad band noise
components were normalized by their total power as measured over a wide
frequency range (see van der Klis 1995a for the specific formulae). The
advantage of this approach is that the integral power is directly fitted to
the power spectra and its error follows directly from a scan in $\chi$$^{2}$
space. 

{\it 2.3: Deadtime Effects.} 

The four-band (I7) data and the high time resolution (T3 and T5) data are
affected by different deadtime processes. In both cases 0.5\% of the incident
counts are lost due to the electronic chain coincidence logic deadtime. Beyond
this, the HTR count rates suffer from only the constant detector deadtime of
5.5\,$\mu$\,sec (Andrews and Stella 1985, but see below). This sort of
deadtime process results in a correlation between successive time bins which
results in a rise in predicted Poisson noise power toward the Nyquist
frequency. In spite of the high count rates in some of our observations, this
frequency dependence of the Poisson noise power spectrum was found to be
always completely negligible. The frequency averaged Poisson noise was
calculated using formula (3.9) of van der Klis (1989) for each individual
power spectrum. When power spectra were averaged the expected Poisson levels
were also averaged with the same weighting. 

The accuracy of the estimated Poisson level is limited by the accuracy to
which the ME detector deadtime is known. Using data from a wide variety of
X-ray sources Berger \& van der Klis (1994) found that for T3 and T5 data, the
high frequency ($\ge$256\,Hz) noise level at a given count rate could be best
fit with a deadtime of 10.6$\pm$0.3\,$\mu$\,sec rather than 5.5\,$\mu$\,sec.
This means that the strength of the HFN is underestimated in our analysis by
0.3--0.4\% (rms). Observations of Cyg~X-3 (Berger \& van der Klis 1994) and a
comparison between the HFN strengths as measured with EXOSAT and Ginga for
various Z sources (Prins \& van der Klis 1997, Berger \& van der Klis 1998)
further suggest that there is extra noise in the EXOSAT ME observations at a
level of $\sim$3.3\% (rms) with a cut-off frequency near 100\,Hz. If this
noise is an instrumental effect of the EXOSAT ME and if it is the same for the
Ar and Xe detectors, then the rms amplitudes of the HFN reported below are
overestimated by about 0.5\% (rms). Note that the two effects approximately
cancel in our case. The effect of these instrumental details on other power
spectral components than the HFN is negligible. 

In the case of the HER data modes (which include HER7, the source of the I7
data) the basic deadtime process is one in which not more than one photon can
be detected in each 1/4096 sec long time-sample. Andrews and Stella (1985)
derived an accurate, within $\leq$0.1\%, expression (their equation 2) for the
count rate correction factor, which we used to correct the count rates in each
channel before calculating the colours. In the cases where we used I7 data for
calculating power spectra, we applied the empirical deadtime corrections
determined by Kuulkers \etal\ (1994,\,1997) to estimate the Poisson level. 

All quoted QPO amplitudes have been corrected for differential deadtime
effects (Lewin, van Paradijs \& van
der Klis; 1988) and for the effects of time binning (van der Klis 1989).
Differential deadtime occurs when the deadtime depends upon count rate
and so the deadtime is larger in the peaks than in the valleys of a signal. 
Hence, the count rate is suppressed by a larger factor in the peaks. The end
result is the rms amplitude goes down. 
The rms amplitudes of the other power spectral components have also been
 corrected for differential deadtime effects but not for the loss of 
sensitivity due to time binning. In the only case where the binning effect 
is not negligible (the HFN), the loss of sensitivity near the Nyquist 
frequency is partially compensated by an unknown amount of power aliased 
from beyond the Nyquist frequency. 

{\large\bf 3: Results} 

{\it 3.1: Overall Behaviour.} 

The colour-colour diagram of Fig.\,1 is made up from the five EXOSAT
observations performed in 1985--86 where Ar four-channel I7 data were
available. The diagram shows a single Z pattern with data from 1985 and 1986 
overlapping on the lower NB. These observations were made under nearly 
identical conditions (e.g. collimator response) and so the primary 
uncertainties in the position of the Z-track from one observation to the next 
are due to changes in the background and detector gain changes.  Background
uncertainties  could result in a 1\% change in the soft colour and a 10\%
change in the hard colour. Using the Crab as a calibrator the gain  changes in
Array Half 1  (the only half used) during 1985 were about 1\% in both the
colours. These uncertainties should be  compared with the shifts seen for
Cyg~X-2; $\sim$3\% in soft colour and $\sim$8\% in hard colour Hasinger
(1987b), Kuulkers \etal\ (1996) and for GX~5-1; 11\% in soft colour and
$\sim$7\% in hard colour Kuulkers \etal\ (1994).

Observations made in 1984 show a  Z track that is slightly shifted in both
colours and with a subtly different shape (see Fig.\,2). The about 5\% drop in
hard colour of the NB/FB vertex visible in Fig.\,2 with respect to Fig\,1 is
within the systematic uncertainty. The 1984 observations were
made under significantly different  conditions from the latter I7 (1985/86)
observation. Between the two sets of  observations detector 3 of array half 1
failed.  Using the Crab as a calibration source and cross checking with
GX\,5-1(Kuulkers \etal\ 1994,  Kuulkers 1995) we found that the loss of
detector 3 and small gain changes between 1984 and 1985 are expected to result
in a 2--3\% increase in the measured soft colour. This is contrary to the
measured $\sim$1.5\% decrease in soft colour between the 1984 and 1985/86
observations.  However there are several other effects that will influence the
soft colour, which can not be adequately calibrated using the Crab. These
effects include the different energy boundaries used for the \sco\
observations, the different energy spectrum of \sco\ as compared to the Crab
and GX\,5-1 and the low collimator responses of the all \sco\	    
observations. Also the collimator responses are larger for 1984
(12\%--23\%) than for 1985/86 (8\%--12\%). These other effects could
explain the differences between the 1984 and the 1985/86 observations and so
there is no evidence for from these data for secular changes in the Z-track 
of \sco\ between 1984
and 1985/86. In the period 1985/86 we can only exclude  only the largest
secular changes seen in GX\,5-1 (Kuulkers \etal\ 1996). 

Because of the differences in their Z tracks, the 1984 and the 1985--86 data
were treated separately. The \Sz\ values were calculated as described in
Sect.\,2.1 independently for the two data sets. The \Sz\ scaling of the
1985-86 data, where both the HB/NB and NB/FB vertices are seen, was also
applied to the 1984 data, where only the NB/FB vertex is present. 

On all hardness-intensity diagrams produced from the 1985-86 (HER7) and 1984
(HER5) data, the minimum hardness (colour) was found to correspond to  minimum
brightness. So, the point \Sz=2, which was set at the minimum in both the soft
and hard colours on the CD should also correspond to minimum brightness.
Fig.\,3 shows this is the case for the 1985--86 data.  A similar variation in
count rate with \Sz\ is found for  the 1984 data. Fig.\,3 clearly shows 
that the count rate is not a good indicator of
\Sz, and by inference the mass accretion rate. 

The count rate versus \Sz\ plots provide a check on the accuracy with which we
can fix the position of the NB/FB vertex upon the CD. We fitted
count-rate/\Sz\ data in the region near the vertex with two intersecting
straight lines. For the 1985--86 data there was no measurable difference
between the position of minimum count rate and the \Sz=2 position.  However,
for the 1984 data the count rate reaches its minimum slightly NB-ward of the
NB/FB vertex, at \Sz=1.978. We have shifted the \Sz\ values of the 1984 data
by $\Delta$\Sz\ = 0.022 to match them with the 1985--86 data. 

This small  discrepancy gives an indication of how well the position of the
NB/FB vertex can be placed. Most of this difference can be accounted for by the 
uncertainty in  determining the placement of the NB/FB vertex rather than from
effects arising from having slightly different energy bands and different
detector responses between the two data sets. Because the HB/NB vertex is more
rounded the uncertainties in its placement are larger. We estimate that the
uncertainties in placing the NB/FB and HB/NB vertices are $\pm$0.013 and 
$\pm$0.05 \Sz\ units respectively. This leads to a uncertainty of $\sim$5\% in
the scaling of the Z-track, The errors on individual \Sz\ values are much
larger being the projection of the statistical errors of the individual colour
measurements (See Fig\,1).

Fig.\,4 shows various representative power spectra in the Xe ({\it filled
circles}) and Ar ({\it open squares}) energy ranges; the drawn lines are model
fits. The spectra are all plotted to the same scale and normalized such that
the power density is in units of (rms/mean)$^2$/Hz (see van  der Klis 1995a
for the formulae). The estimated Poisson level has been subtracted, and the
spectra have been corrected for differential deadtime effects (van der Klis
1989b), but not for time binning and aliasing. The segment of the Z track
corresponding to each power spectrum has been indicated in Fig.\,1. 

In broad terms, the features of these power spectra are similar to those noted
by HPM89 and Hertz \etal\ (1992). The rms amplitude of the VLFN increases
along the Z track (Fig.\,4A to F; note the layout of Fig.\,4; it reflects the
shape of the NB-FB part of the Z track), while that of the HFN decreases. No
QPO are observed in the HB (the 90\% confidence level upper limit in the Xe
band is 3\%~rms), but the peaked LFN is obvious (Fig.\,4A). Detailed analysis
shows that the LFN persists into the upper NB. The LFN is just visible in
Fig.\,4B. At positions just before the QPO appear (\Sz\ = 1.5; not shown  as a
separate plot in Fig.\,2) only VLFN and HFN are detected. The high time
resolution and high count rate of the data make it possible to trace the HFN
out to high frequencies. In the lower NB (Fig.\,4C) the well known 6\,Hz QPO
(the NBO) is evident. Upon entering the FB (Fig.\,4D) the QPO frequency
increases. The highest FBO frequency measured is 21\,Hz. Further along the FB
(Fig.\,4E) there is some evidence for an additional peaked component near
5.5\,Hz. Higher still on the FB (Fig.\,4F) there is only strong, steep, VLFN
and relatively weak HFN.

A comparison between the Ar and Xe power spectra clearly shows the strong
energy dependence of the various components. The fractional rms amplitudes of
all components are larger, by up to a factor of 3 (9 in power), in the higher
Xe energy band. The strong energy dependence of the FBO strength we report
here is a new result. In Fig.\,4D there is only a slight indication of any FBO
(15\,Hz) in the Ar data, while they are easily visible in the Xe data; the
difference in fractional rms amplitude is a factor of 2.5. Table\,2 gives the
fractional rms amplitudes of the various components for both the Ar and Xe
power spectra of Fig.\,4. 

{\it 3.2: The broad band noise components: VLFN, LFN and HFN.} 

In Fig.\,5, we present the best fit parameters of the VLFN, LFN and HFN
obtained from the 128-sec Xe power spectra selected as described in
Sect.\,2.2. Two sets of fits were performed; one with the HFN slope
($\alpha_{HFN}$) fixed at zero, the other with this parameter free. In most
fits where $\alpha_{HFN}$ was a free parameter, it is slightly positive, but
consistent with being zero at the 2$\sigma$ level. When $\alpha$$_{HFN}$ was
fixed to zero the $\chi$$^{2}$ value increased, but usually not significantly,
nor did the other fit parameter values change significantly. For ease of
comparison with other studies, the values reported in Fig.\,5, except those of
$\alpha$$_{HFN}$, are those obtained from the fits with
$\alpha$$_{HFN}\equiv,0$. 

The VLFN rms amplitude integrated over 0.001--1\,Hz increases monotonically
from 1.0\% in the HB to 25\% in the upper FB (the 2$\sigma$ lower limit there
is 18\%). The VLFN is stronger in the Xe band than in the Ar band on the NB
and FB, but on the HB the difference is small (see Table\,2 and Fig.\,4).
However, as the VLFN is very weak on the HB and only evident in the one or two
lowest frequency points, the determination of the VLFN parameters there is
uncertain. The VLFN power law index ($\alpha$$_{VLFN}$) is lowest on the HB,
increases from 1.2 to 1.7 along the NB, and is constant on the FB. The mean FB
value is 1.76$\pm$0.04,  practically the same as that measured by Hertz \etal\
(1992), 1.72$\pm$0.01. Several of the individual power law index measures,
especially along the FB, are close to 2. There may be some low frequency
leakage (Deeter 1984) from the 5-35~min duration, 10-200\% flares on the FB. 

The LFN appears as ``peaked noise" in the HB. The amplitude in Xe is highest
(3.3\%\,rms) at the lowest observed \Sz, drops when \sco\ enters the upper NB,
and further along the Z track (\Sz$\ge$1.5) it is undetectable with a 97.5\%
confidence upper limit of 2\%\,rms. Both the LFN slope and cut-off frequency
change rapidly near the HB-NB vertex. The slope, $\alpha_{LFN}$, changes from
$-1.5$ to about 0.0 there, while $\nu_{cut}$ increases from about 1.4 to
$>$7\,Hz. These steep dependencies on \Sz\ reflect a broadening of the LFN
shape. 

The HFN amplitude (integrated over 1--500\,Hz) gradually decreases from the HB
(8.4\%) to the FB (5.5\%\,rms). The weighted average of all measurements of
the power law slope of the HFN is 0.066$\pm$0.001, indicating an on average
slightly ``red" noise shape for this component. The apparent dip in
$\alpha_{HFN}$ near the NB/FB vertex is not very significant: the average
$\alpha_{HFN}$ in the range $1.8<$\Sz$<2.2$ is only 2$\sigma$ less than the
overall average. We do find two significant features in the variation of the
HFN cut-off frequency with \Sz. The cut-off frequency drops from \about 70 on
the HB to
$\sim$35\,Hz (4.4\,$\sigma$ difference) on the NB, and then increases along
the FB to about 100\,Hz (95\% confidence). This last trend seems to be at odds
with the results of Hertz \etal\ (1992), who found that the width of the half
Gaussian they used to fit the HFN decreased as \sco\ moved up the FB. 

The amplitude of the HFN is, at similar positions on the Z track, consistently
$\sim$4\% rms higher than that measured by Hertz \etal\ (1992) using Ginga.
This difference can be explained by the difference in energy response of the
two detectors. The EXOSAT Xe detectors are more sensitive to higher energies,
where the HFN is stronger, than the Ginga LAC. A similar difference in
amplitude is found for the VLFN component. 

The HFN is detectable above the noise out to about 300\,Hz in the upper NB and
to at least 100\,Hz in the FB (Figs.\,4A, B and E). In these regions the
source count rate is highest, making the HFN more readily detectable against
the Poisson noise level. As discussed in Sect.\,2.3, uncertainties in the
instrumental effects on the power spectrum make these estimates uncertain. 

{\it 3.3: The NB-FB QPO.} 

The results of the QPO fits to power spectra of the 16 and 32-sec data
segments averaged together over intervals of constant count rate or \Sz\ as
described in Sect.\,2.2 are displayed in Figs.\,6\,\&\,7. 

QPO are detected ($\ge$2.0\%\,rms) over only a small part of the Z track (\Sz\
$=$\,1.5--2.2), or only \about17\% of the total \Sz\ range. NBO occur on only
the lowest one third of the NB while FBO occupy only the lowest 10\% of the
FB. The QPO are weaker in Ar than in Xe on both the normal and flaring
branches. 

Previous studies have shown that as the QPO frequency increases the width
increases in step keeping the relative width roughly constant. The relative
width varies is the inverse of the QPO coherence. By binning in \Sz\ (Fig.\,6)
we reveal that there are systematic changes in the relative width
($\Delta\nu/\nu$) along the Z-track.  It is lowest ($\sim$0.27; highest QPO
coherence) in the lower NB (\Sz=1.75), and increases on the FB (to $\sim$0.41
at \Sz=2.05) and possibly in the upper NB. The coherence time-scale
($\tau_{coh}\equiv1/\pi\Delta\nu$) ranges from $\sim$0.18\,sec (1 QPO cycle at
6\,Hz) in the NB to $\sim$0.05\,sec (0.75 QPO cycles at 15\,Hz) in the FB. 

In Fig.\,7, showing the results of individual fits to each data section with
QPO, the symbols refer to same observations as those in Fig.\,1.  Figs.\,7a
and 7b show the dependence of NBO/FBO fractional rms amplitude on \Sz\ using
the combined Ar+Xe (1984) and Xe only (1985--86) data. The QPO are weaker in
the lower energy Ar band than in the Xe band on both the normal and flaring
branches. In both energy bands the QPO amplitude increases with \Sz. In the Xe
only data, which spans a greater range of \Sz, a clear increase in fractional
rms amplitude is seen along the NB, from $\sim$3\% at \Sz=1.55, to $\sim$5\%
at \Sz=1.95, and up to 8\% in the FB. In the lower energy Ar+Xe data, the QPO
fractional rms is about 2.5\% in the lower NB and between 2 and 4\%, but
usually $>$3\% in the FB.  In the higher count rate Ar+Xe data there is a
steep increase in QPO rms amplitude (1\%\,rms within 0.05 \Sz\ units) at the
NB/FB vertex. There is possibly a similar steep increase  (2\%\,rms within
0.05 \Sz-units) of the QPO amplitude in the Xe band data. But in view of the
large scatter in measured rms amplitudes on the FB, the identification of
steep changes in rms amplitude at the NB/FB vertex is only tentative. 

Clearly there is a dramatic jump in the QPO frequency at the transition from
the NB to the FB (Fig.\,7c). The actual transition is not resolved. It occurs
within a span of $\leq$1.5\% of the Z track, or $\leq$8\% of the span over
which QPO are detected.  The QPO frequency is between 5.5 and 7\,Hz all the
way along the NB from \Sz=1.5 to \Sz=1.9. At \Sz=1.95 the QPO frequencies are
between 6.4 and 8.5\,Hz, and at \Sz=1.98 between 6.5 and 12\,Hz. The average
frequency is $\sim$8\,Hz here. Upon entering the FB (\Sz$\ge$2) the average
QPO frequency jumps from  $\sim$8\,Hz to $\sim$12\,Hz, and then increases more
gradually with \Sz\ and brightness. The highest frequency measured
($\sim$\,21\,Hz) was during a flare in the Feb\,1985 data. 

A frequency-intensity diagram of the 1985-86 Xe, data sets is shown in
Fig.\,7d. For the Xe+Ar data this diagram looks qualitatively the same but at
higher count rates. The same features as in the frequency versus \Sz\ diagram
are visible. By also using high time resolution count rate data for which there
are no simultaneous energy resolved data,
(included in Fig\,7d but not Fig\,7c) 
we find that the average frequency
increases slightly but significantly along the upper NB as \sco\ approaches
the NB/FB vertex.  As the count rate changes from 3000 to 2450
cts/sec/8~detectors, corresponding (using the count rate vs \Sz\ relation) to
\Sz=1.57 and 1.87, respectively, the average QPO frequency increases from
5.96$\pm$0.04\,Hz to 6.30$\pm$0.03\,Hz. 

On both the frequency-\Sz\ and frequency-intensity diagrams there is more
scatter on the lower NB near the transition and on the FB than on the upper
NB. This is evident in the dynamical power spectra (16 or 32\,sec time
resolution), which show the band of NBO powers becoming broader and more
diffuse near the NB/FB transition. Near the NB/FB vertex the QPO frequency is
more sensitive to \Sz\ than in other places on the NB, while on the FB the QPO
frequency is changing on the same time-scale as the flares, i.e., minutes to
hour(s). In both cases the time-scale of the changes in QPO frequency becomes
comparable or shorter than the typical integration times ($\ge$200\,sec) used
to fit the QPO and measure the position on the Z track. The \Sz\ values become
difficult to measure on time scales $<$200\,sec. Some of the scatter on the
frequency-\Sz\ diagram can therefore be explained by the increased difficulty
in measuring \Sz, but this does not hold for the scatter in the
frequency-intensity diagram. Here the extra scatter could be intrinsic to
\sco, indicating either the presence of hysteresis or the influence of a
another variable besides {\mdot} on QPO frequency and/or brightness. Direct
tests for hysteresis are made in Sect.\,3.6 and a careful examination of the
QPO behaviour during normal/flaring branch transitions (Sects.\,3.4 and 3.5)
finds evidence for the effects of an additional variable. 

{\it 3.4: Normal/Flaring Branch Transitions.} 

Upon examination of Fig.\,7 and the dynamical power spectra we found that
there are three QPO frequencies that are important in the description of the
different modes of behaviour of \sco. At frequencies $<$7\,Hz \sco\ is on the
NB and there are only weak correlations between \Sz, QPO frequency and 
count rate. Overall as \Sz\ increases the QPO frequency increases and the
count rate decreases.  Between 7 and 8\,Hz \sco\ is on the lowest part of the
NB (\Sz$>$1.9) and there is a positive correlation of frequency with \Sz, but
an inverse correlation with count rate. As \sco\ makes the transition between
the NB and FB the QPO frequency changes from 8 to 12\,Hz. Here the relation
between frequency and position cannot be exactly determined  with the current
data. Above $\sim$12\,Hz \sco\ is on the FB, where there is a strong positive 
correlation between QPO properties, \Sz\ and brightness. We use these
three  QPO frequencies (7,\,8,\,12\,Hz) to discuss the different time scales 
on which the QPO frequency changes near the NB/FB vertex. 

In the following discussion of the QPO behaviour keep in mind that we could
not reliably measure the QPO frequency on time scales less than
$\sim$200\,sec. By also considering the count rate behaviour we can sometimes
deduce that the QPO frequency must have changed rapidly between two adjacent
sections of data. The shortest time-scale measured in this way for a
significant change in QPO frequency is 90\,sec. We cannot make any general
statements on frequency variations faster than this. In most cases we found
that the changes in QPO frequency took the same time no matter in 
which direction
\sco\ was traveling on the Z track. 

\newcounter{bean} Based upon a careful examination of the dynamic power
spectra, intensities and colours we find that one can distinguish four
patterns of behaviour near the NB/FB vertex. 

\begin{list}{\Roman{bean}}{\usecounter{bean}\setlength{\rightmargin}
{\leftmargin}}

\item {\bf\it Grand Transitions.} These are very obvious steady transitions
between the NB and the FB or vice-versa. This is the sort of transitions
described by Priedhorsky \etal\ (1986). These transitions separate the 
historical active
and quiescent states which correspond to the FB and NB respectively. On the FB
side of a {\em grand transition} \sco\ shows classic flaring behaviour. During
flares, count rate changes of up to a factor of two occur on time scales of
minutes to tens of minutes. Each flare represents a movement up and down the
FB with associated changes in brightness, colour(s), QPO frequency and other
power spectral parameters. On the NB side there are no flares, but slower,
tens of minute to hours time-scale changes in \Sz. Here the brightness changes
are less than a factor of 2 and the QPO frequency varies between 5.5 and
8\,Hz. Within the EXOSAT data set we have found 5 {\it grand transitions}
which are indicated in the ``Branch" column of Table\,1. During these
transitions  \sco\ moves continuously and smoothly (i.e., with no reversals in
the direction) in the CD, HID and frequency-intensity diagram. The change in
QPO frequency involved in a {\it grand transition}, i.e., from $\sim$7\,Hz to 
$>$12\,Hz or vice-versa, takes $\sim$600--1000\,sec. The speed with which the
frequency changes is different over different frequency intervals. The change
from 8\,Hz to $>$12\,Hz or vice-versa is most rapid, taking between
$<$120\,sec and 300\,sec, while the frequency change between 7 and 8\,Hz
usually takes $\sim$300 to $\sim$700\,sec. On either side of the transition
the frequency changes are more gradual. On the NB frequency changes in the
range 5.5 to 7\,Hz take typically many hundreds of seconds while on the FB the
time-scale of the frequency changes is the same as that of the flares. 

\item {\bf\it FB dips.} On the FB, during relatively brief ($<$200--600\,sec)
periods of lower count rates, FB QPO appear. This is the sort of behaviour that
was noted by van der Klis \etal\ (1987). We have found tens of these dips in
several observations. In all cases the frequency never dropped below about
10\,Hz, so there are no switches to NBO. The change in frequency between 15
and 10\,Hz can take less than 120\,sec. 

\item {\bf\it NB dips.} On the NB, there are also relatively brief 
($\sim$600\,sec) dips in count rate as the QPO frequency increases to 8-9\,Hz
 before
returning to the norm of 5.5-6.5\,Hz. \sco\ does not enter the FB proper
during these episodes. The {\it NB dips} can be thought of a sub-group of the
{\it rapid excursions}.

\item {\bf\it Rapid Excursions.} These are rapid changes in frequency from
$\sim$7\,Hz to $>$12\,Hz, i.e. well into the FB, and back.  We found 8 rapid
excursions in the entire high time resolution data set, i.e., 3 Aug 1984 (2),
25 Feb 1985 (2), 25 Aug 1985 (1), 27 Aug 1985 (3). They all started and ended
in the NB. Usually the change from 7 to $>$12\,Hz took $\sim$120\,sec, but in
one case it took less than 90\,sec. The return to the NB (from 8 to
$\sim$\,6.5\,Hz) is generally slower; 300--600\,sec. This sort of behaviour
has not been previously described. 

\end{list} 

When \sco\ moves through the NB/FB vertex, the count rates and colours reach a
minimum when the QPO frequency is near 8\,Hz. These conditions define the
NB/FB vertex. Thus, in an excursion from NB to FB and back the count rates and
colours should pass through two equal depth minima, each time as the frequency
passes 8~Hz. In between these two minima, the count rates and colours reach a
maximum as the frequency reaches its peak value.  Just before making a
transition, i.e., when the QPO frequency is near 7.5\,Hz the Xe count rate,
soft and hard colours are 2220--2360 counts/second/8~detectors,
0.903$\pm$0.007 and 0.661$\pm$0.009 respectively (all  uncertainties in this
paragraph correspond to the observed total scatter).  During the transition,
the hard colour is more sensitively dependent on \Sz\ on the NB side of the
vertex, while the soft colour is more sensitive on the FB side of the vertex.
At the vertex the minima are located in Ar count rate at 4108$\pm$112
counts/second/8~detectors (at 10\% response; 1986 observations; see Sect.\,2
for the issues to do with the Ar collimator response), in Xe count rate at
2247$\pm$82 counts/second/8~detectors, in soft colour at 0.907$\pm$0.015, and
in hard colour at 0.650$\pm$0.015. On the FB side the count rate, soft and hard
colours increase to 2230--2660 counts/second/8~detectors, 0.91$\pm$0.01 and
0.663$\pm$0.005, respectively, when the QPO frequency reaches 15.5\,Hz. 

Examining the best-covered transitions on timescales ranging from 2--32\,sec 
we find, using a $\chi$$^{2}$ test for constancy, that even within individual
observations there are variations in the depth of the count rate   minima
during NB/FB transitions  beyond those expected from counting statistics (95\%
confidence). The full  range in the depth variation is $\sim$1\% of the count
rate at the NB/FB vertex. By inference the colours at the NB/FB may also vary.
Thus the vertex covers a range in brightness and possibly colours, with \sco\
traversing slightly different paths in the CD or HID as it moves between
branches. In other words; the one-to-one correspondence between Z track
position, measured by \Sz\, and other variables breaks down near the NB/FB
vertex, suggesting that there is another parameter, in addition to {\mdot},
governing the behaviour of \sco. 

We found that in two of the {\it rapid excursions} \sco\ did {\em not} follow
the standard behaviour when the QPO frequency went through an excursion from
levels appropriate to the NB to higher values. These {\it anomalous frequency
excursions} are discussed next. 

{\it 3.5: Anomalous Frequency Excursions.} 

Two cases of anomalous frequency excursions were found during a careful
examination of the entire high time resolution EXOSAT data set. The first
example discovered, occurred on 13 Mar (day 72) 1986, is apparent in
Fig.\,7b,c,d, where two points ({\it circled}) on the NB in the CD and HID
have higher frequencies (8.1\,$\pm$\,0.3\,Hz and 11.1\,$\pm$\,0.5\,Hz.) and
amplitudes (Xe band) higher than is usual for the NB. These frequencies and
amplitudes are more appropriate to the FB, but the corresponding values of
\Sz, near 1.9, place these points squarely on the NB. The dynamical power
spectrum shows a fast increase and decrease in frequency with a total duration
of $\sim$\,5\,min. This sort of frequency behaviour is very similar to that of
{\it rapid excursions}, but there is no change in the colours and consequently
the derived \Sz\ values. The other instance of an anomalous  frequency
excursion is the looping behaviour first noted by van der Klis \etal\ (1987)
on 30 Aug (day 242) 1983. Here, \sco\ executed a loop upon the
frequency-intensity diagram, initially passing from the NB to FB through the
NB/FB vertex but then  returning {\bf directly} to the NB, by-passing the
vertex. Once again the dynamical power spectrum is very similar to that of a
{\it rapid excursion} but the hardness and count rate do not show a peak,
bracketed by two dips as the QPO passes through 8\,Hz. The count rates,
frequency, and colour behaviour during these two events was scrutinized in
greater detail and compared to what is found during standard NB/FB
transitions. 

The time-wise behaviour of \sco\ during the March 1986 frequency excursion is
summarized in Fig.\,8. Before the start time of this figure \sco\ was in the
mid-NB (i.e., with lower \Sz\ values, higher colours and intensities than
shown) and generally moving down the NB. Just after $1.5\times10^4$\,sec we
see that the QPO frequency reaches a maximum value of $\sim$16\,Hz. This point
is labeled ``Peak" in Fig.\,8. At the same time the count rates from the Ar
and Xe detectors reach a maximum. At this point the QPO frequency and count
rates are typical of the FB. About 270\,sec before, and $\sim$1200\,sec after
the peak in count rate and QPO frequency there are two dips in the counting
rate, at the same time as the QPO frequency passes through 8\,Hz. Both dips
are labeled as such in Fig.\,8. 

At the time of the count rate dips there are two marginal drops in the colours;
see top two panels of Fig.\,8. As discussed above, the changes in the colours
are expected to be small, with the change from minimum to the values
corresponding to 16\,Hz QPO being 0.03\,$\pm$\,0.025 (full range) for the soft
colour and 0.013\,$\pm$\,0.02 (full range) for the hard colour. Changes of
this size might be present in Fig.\,8. At the first dip the \Sz\ values
increase reflecting the small decrease in both colors, but then, instead of
rising to values $>$2 as would be appropriate for a rapid change to the FB,
\Sz\ drops again. At this time the colours do not vary in unison which
explains  the $<$2 values for \Sz. The values of \Sz\, at this time, were not
modified during the search and replace procedure used for wayward \Sz\ points
(Sect.\,2.1).  To show the changes in colour more clearly we calculated a
hardness ratio using the Ar and Xe band counting rates. The resulting hardness
ratio (colour) is plotted in the middle of Fig.\,8. There are clearly two dips
in the hardness ratio flanking a general rise in hardness but the Ar/Xe colour
does not track the count rates and QPO frequency closely as expected. 

The observed pattern, a peak in count rate, QPO frequency and, marginally,
colours bracketed by two dips associated with 8\,Hz QPO, is  just what is seen
during {\it rapid excursions}. What is not normal during this frequency
excursion is the behaviour as the source moves to and from the peak in
count rate and frequency as seen from the trajectories traversed on
frequency-intensity, frequency colour(s) and colour-colour diagrams, which we
discuss next. 

The frequency-intensity diagram of Fig.\,9 is similar to that of Fig.\,7d but
with higher time resolution during the frequency excursion, shows how the
count rate varies before and after the peak QPO frequency.  The 1986 frequency
excursion is displayed as a dotted line, beginning at the point labeled
`Start`, then moves via the `1st Dip` to the 'Peak', then returning to near
the NB/FB vertex at the `2nd Dip` just before reaching the `End` point, after
which \sco\ remains on the NB. Clearly, in this diagram \sco\ does not follow
the usual path between the normal and flaring branches. Both count rate dips
are at comparable intensities to other observations when \sco\ was near the
NB/FB vertex, but the subsequent movement to and from the peak in count rate
and frequency  seems to occur at an count rate that is
$\sim$200~cts/sec/8~detectors higher than the standard path on a
frequency-intensity diagram.  This difference in count rate would correspond
to an 8\% difference in collimator response, or a 16 arcmin  pointing offset
between the 1986 and 1985 observation.  We therefore checked if there was any
motion in the pointing direction of EXOSAT during this observation or a
difference in collimator response with respect to other observations. An
examination of the house-keeping data showed stable pointing, and checking the
original observational logs kept by ESA established that the Xe detectors were
in all observations pointed directly at \sco. Two hundred counts is far
greater than the full range ($\sim$26 cts/sec/8~detectors) of the variations
found in the background counting rates using observations spanning the years
1984--86. Examining the veto and  guard count rates there is no indication of
any change in the background at the time of the frequency excursion. The gain
of the Xe detectors seems stable, since in all cases except the anomalous
frequency excursions the Xe count rate at a particular \Sz\ is reproducible
over the lifetime of EXOSAT. We conclude that the different path taken in the
frequency-intensity diagram during this excursion is intrinsic to \sco. 

This conclusion is supported by the frequency-colour diagrams. On both
possible frequency-colour diagrams (Figs.\,10a,b) the path of \sco\ is
slightly shifted with respect to the other points in a similar way as on the
frequency-intensity diagram. Such a shift cannot be caused by changes in the
collimator response. 

The motion on the colour-colour diagram of \sco\ during the frequency
excursion is shown in Fig.\,11. The position of the `1st Dip' and `2nd Dip'
are marked by a filled circle and square respectively while the `Peak' is
marked with a filled triangle. The colour-colour points associated with the
high QPO frequencies are surrounded on all sides and even coincide with points
associated with 6\,Hz QPO. At no time does \sco\ apparently enter the flaring
branch. However, it should be noted that during the 1986 observations there
were no standard transitions to the FB and so the FB data of Fig.\,11 come
from a year earlier. Therefore the frequency excursion could represent motion
into a FB that is on top of the NB in the colour-colour diagram. We consider
this possibility unlikely because the  colour-colour diagrams appear stable
from year to year. Observations  well separated in time show that the
positions of the NB and FB in the colour-colour diagram are reproducible, also
the separation between the two branches shown in Fig.\,1 using 1985-86 data is
similar to that found in 1984 (see Fig.\,2). 

In summary, there are two possible ways to interpret what happened; \sco\ made
a rapid excursion along a FB that is in an unusual place on the CD, or it
remained on the NB while the QPO frequency changed briefly to values
appropriate for the FB. 

The second anomalous frequency excursion is seen as a loop on the
frequency-intensity diagram (van der Klis \etal\, 1987) during the
observations on 29/30 Aug (day 241/242) in 1983. The dynamic power spectrum is
similar in appearance to a rapid frequency excursion. At the beginning of this
observation the EXOSAT satellite pointing direction was moving. Although this
movement apparently ended well before the unusual NB-FB-NB transition, the
aspect information does not unequivocally show the pointing as being stable.
So, it is possible (but in our experience unlikely) that the intensities are
not  entirely reliable.  On this occasion the multi-channel data from the ME 
detector were affected by data overflows. Two-channel GSPC (Gas Scintillation
Proportional Counter) data are  available in the EXOSAT data base maintained
at ESTEC and GSFC. The 5--8 and 8--15\,keV light curves were extracted and a
GSPC colour (hardness ratio) curve was calculated. The ME and GSPC data could
be synchronized because the start times are recorded with better than 1\,sec
accuracy. 
  
The time-wise behaviour during the loop is summarized in Fig.\,12. Initial
observations, ending 200\,sec prior to the data of Fig.\,12, show \sco\ on the
FB. \sco\ was generally moving toward the NB/FB vertex.  Judging by the
colours and the QPO frequencies \sco\ was near the NB/FB vertex in the early
part of the observation when the pointing was not stable. In fact \sco\ spends
a long time (2000--3000\,sec as opposed to the normal $\sim$1000\,sec total
duration for most transitions) near the vertex with a QPO frequency being
between 8 and 12\,Hz. As the frequency increases reaching a maximum of
$\sim$18\,Hz (labeled F$_P$ in Fig.\,12) there is an increase in the count
rates (ME and GSPC) but no increase in colour, contrary to standard transition
behaviour. This places \sco\ outside the FB on the frequency-colour diagram as
delineated by the earlier measurements on the same day. As the QPO frequency
decreases to near 16\,Hz the count rates continue to increase to a maximum
(labeled as I$_P$). At this time the colours are increasing. Soon after the
peak in count rate the frequency drops from 16 to 9 Hz (labeled Jump). During
this abrupt drop in frequency  there is no indication of a drop in either ME
or GSPC count rates or colour as would be expected for a passage through the
NB/FB vertex. Any such passage would have to be have been very rapid, say,
$<$30\,sec to go undetected in the light curves. The times most appropriate to
be designated count rate dips are labeled A \& B on either side of the
frequency excursion. At A the colour is low but does not increase as the
frequency becomes $>$8\,Hz. At time B there is no indication of a change in
colour. After the frequency jump, \sco\ moves toward the NB/FB vertex. At the
very end of Fig.\,12 there is a drop in colour and count rates which may
indicate \sco\ approaching the NB/FB vertex. During the next ME observation,
some 160\,sec later (no GSPC data for 10660\,sec) \sco\ is found in the NB
with stable 6\,Hz QPO. 

Fig.\,13 shows the colour-frequency diagram for \sco\ on 29/30 Aug 1983. Data
from the three observation periods, separated by short (200 and 160\,sec)
intervals, are displayed with different symbols. In the first observation
({\it filled triangle}) \sco\ was on the FB. The third observation ({\it
filled square}) finds \sco\ on the NB.  During the middle 
observation ({\it filled circle}), covered in Fig.\,12, \sco\ executes a large
loop on the colour-frequency diagram. The positions of the maximum frequency 
(F$_P$) and count rate (I$_P$) as well as the positions of the count rate dips
(A \& B) are marked. Only point A is near the vertex. The jump between the FB and NB
frequencies occurs within 120 sec with no indication of any drop in the colour
or count rate that would indicate a passage through the NB/FB vertex. 

We have checked the veto and guard rates and find no evidence for any change
in the background at the time of the frequency excursion. Therefore we can
conclude that \sco\ made either a very quick ($<$30\,sec) transition from the
mid-FB to the mid-NB, passing through the vertex too rapidly to be measured,
or jumped directly between the flaring and normal branches. 

The two anomalous frequency excursions presented in this section and the
different brightness levels during NB/FB transitions discussed near the end of
Sect.\,3.4 represent the first evidence that a strict one-to-one relationship
between spectral state and QPO behaviour does not hold on short (several
100\,sec) time scales.

{\it 3.6: Movement along the Z track.} 

Using the \Sz\ parametrization it is possible to investigate the kinematics of
the motion along the Z track. The simplest measures, are the velocity and
speed of motion. The velocity, \Vz(i) at time $T(i)$, was defined as 

 \Vz(i) = [\Sz(i+1) -- \Sz(i--1)]/[T(i+1) -- T(i-1)] 

where \Sz(i) is the position at time T(i). 

We calculated the velocities as a function of \Sz\ on two time scales: 96 sec
(3$\times$32\,sec position points) and 588\,sec (3$\times$196\,sec). We found,
as did Hertz \etal\ (1992), that the range of possible velocities, and the
variance in the velocities, increased up the FB. In addition, a similar, but
much less well-pronounced trend was found in the 96\,sec measurements as 
\sco\ moved on the NB. However, away from the FB, on a time scale of 588\,sec
the variance in \Vz\ is indistinguishable from a monotonic function of \Sz,
being lowest in the HB and highest in the FB. These trends in \Vz\ were found
to exist at the 95\% confidence level using Bartlett's test for many variances
(Walpole and Myers 1978). The distribution in velocities was symmetric about
zero.
Specifically, on the FB this means that during flares the time spent on the
rise and fall are on average the same. 

The velocity can therefore be meaningfully recast as the speed along the Z
track.  The speed is defined as the  absolute value of the velocity; it has no
information about the direction of motion along the Z track. Fig.\,14 is a
plot of the average speed versus position. It can be clearly seen that on
96\,sec time scales \sco\ moves most slowly just before the NB/FB vertex,
i.e., where the NBO frequency is between 7 and 8\,Hz. As the source moves
through the vertex into the FB the average speed increases rapidly. Thus one
expects that also the N/FBO frequency changes more rapidly in time on the FB.
This accords with the rates at which the QPO frequency was found to change
near the vertex: slowly on the NB side and much faster on the FB. The average
speed increases along the flaring branch which could indicate that larger
flares are steeper (higher amplitude but similar duration) or the flares are
convex in shape. 

The acceleration 

 \Az(i) = [\Vz(i+1) -- \Vz(i--1)]/[T(i+1) -- T(i--1)] 

was also found to have a larger scatter in the FB than in the HB-NB. No
meaningful correlations were found between acceleration and position or
velocity. 

A direct search was made for hysteresis in the dependence of power spectral
shape on velocity. Each observation was split into separate \Sz\ ranges. Power
spectra were sorted according to the sign of the velocity. Two weighted
average power spectra were formed; one where the velocity was positive and the
other where it was negative. The average powers in a number of frequency
ranges (0.01--1\,Hz, 1--10\,Hz, 10--20\,Hz, 20--40\,Hz, 40--60\,Hz) were
compared using the Student's t-test with unequal variances. The variances of
the powers in each frequency range are unequal because there are generally
slightly different numbers of positive and negative velocity power spectra. In
all cases sufficient power spectra were averaged so that the average powers
were very close to being normally distributed. In two cases there was a
significant (95\% confidence) difference in average power within the
10--20\,Hz frequency band. Both cases were near the NB/FB vertex. This is just
where hysteresis could be accommodated given the increased scatter in QPO
frequencies near the vertex. However, there is a more mundane explanation.
Since, near the vertex, the QPO frequency depends sensitively upon the
position, any difference in the distribution of \Sz\ values corresponding to
the positive and negative velocity power spectra will lead to different
average QPO shapes. This explanation was tested as follows. 

There were different numbers of individual power spectra averaged together in
the positive/negative velocity average power spectra. The average spectrum
with the greater number was recalculated but with some individual spectra
removed, such that number of individual power spectra in each average spectrum
was made equal. Either the individual power spectra corresponding to the
highest speeds or those corresponding to the lowest speeds were removed. In
either case (more positive or more negative velocities), for all \Sz\ regions,
the statistical tests now showed no significant difference between the
positive and negative velocity average power spectra. In addition all pairs of
original average power spectra were fitted with a VLFN + HFN components and
where necessary LFN or Gaussian peak(s). The fit parameters for all pairs of
power spectra were the same within their 2 $\sigma$ errors. Our conclusion is
that there is no evidence for hysteresis in the power spectra. 

We made another search for hysteresis by testing if \sco\ took different paths
on the CD depending on whether it was traveling up or down a branch. In
several \Sz\ ranges spanning the Z track we tested if there was a difference
in the average deviation from the smooth Z track (D$_Z$) depending on whether
the velocity was positive or negative. We found no significant difference
(95\% confidence) in the average \Dz\ in any \Sz\ range. 

{\large\bf 4: Discussion.} 

We have presented a comprehensive description of the variation, as a function
of Z-track position \Sz, of all power spectral components of \sco. The Z-track
position is considered to be a measure of the mass transfer rate onto the
compact object for reasons outlined in the introduction. All the power
spectral properties are found to vary in a systematic, but not always smooth
fashion with \Sz. The QPO frequency and the HFN cut-off frequency change
abruptly at the normal/flaring branch vertex. Here we compare our results with
those found for other sources and with the current theoretical understanding
of QPO. 

For any comparison between Z sources one must bear several limitations in
mind. Three other sources, namely GX~5-1 (Lewin \etal\ 1992, Kuulkers \etal\
1994), GX~17+2 (Kuulkers \etal\ 1996) and GX~340 (Kuulkers \&\ van der Klis
1996), have now been examined in sufficient detail to accurately characterize
the variations in all power spectral parameters as a function of \Sz. Cyg~X-2
is the best characterized of the other Z sources. A direct comparison of our
results can be made with those of Kuulkers \etal\ (1994, 1997) because of the
common definition of \Sz. We shall also compare our results with those of
Hasinger \etal\ (1990) for Cyg~X-2 and Lewin \etal\ (1992) for GX~5-1, in
which a slightly different scaling was used, in that the lowest observed
count rate end of the HB was fixed at rank=0, the HB/NB vertex at 1 and the
lowest brightness point observed on the NB, as opposed to the NB/FB vertex,
was fixed at rank=2. The thus defined ``rank number" is therefore not scaled
by the length of the NB as is \Sz. A direct comparison of \Sz\ with rank
number can only be reliably made on and near the NB; otherwise only general
trends with position can be compared. Some earlier studies of other Z-sources
can be recast in terms of \Sz\ because the power spectra are sorted according
to their place upon a CD or HID. This is the case for the studies of GX~340+0
by Penninx \etal\ (1991) and to a lesser extent GX~17+2 by Penninx \etal\
(1990). 

A similar situation exists when trying to compare our results with other
studies of \sco. The concept of using an arc length  to measure the position
along the Z track originated with Hertz \etal\ (1992). Their parameter s$_z$
is an arc length along the Z track with the NB/FB as the zero point. It is
specific to the colours used. However, we can make valid comparisons with the
Hertz \etal\ work because of the good coverage of the FB. We can fix,
approximately, three points that are at the same position of the Z track and so
approximately rescale the arc lengths of Hertz \etal\ to our \Sz\ parameter.
The three calibration points are: the NB/FB vertex, the position at the top of
the flaring branch (assuming both studies cover the entire FB) and the point
where the FBO become undetectable. 

Other complications are that various authors have used different functional
forms to represent the power spectral components and different integration
ranges for measuring the strength of the broad band noise components. The
strengths of all components have a strong energy dependence, so there are also
differences in measured amplitudes depending upon the energy band and detector
used. There are also differences in the way in which various authors take into
account deadtime and background. 

Bearing in mind these comments we will give a summary of our results and a
comparison with similar results for \sco\ and other Z-sources. 

{\it 4.1: VLFN.} 

It has been suggested (van der Klis 1991) that the VLFN is due to the motion
of the source along the Z track. This idea is consistent with the spectral
calculations of Psaltis \etal\ (1996). The similarity we find in the behaviour 
as a function of \Sz, of the VLFN amplitude and the average speed of motion
along the Z track, supports this contention. Both are approximately constant
along the HB-NB and then increase rapidly along the FB. It is possible to
quantitatively test this hypothesis by comparing the rms variability spectrum
of the VLFN and the slope of the Z track at the same \Sz\ on either a
hardness-intensity diagram or a colour-colour diagram. In a CD with soft
colour C$_S$ and hard colour C$_H$, we can consider two colour-colour points
($I_2$/$I_1$,$I_4$/$I_3$) and
(($I_2+a_2$)/($I_1+a_1$),($I_4+a_4$)/($I_3+a_3$)) as being representative of
the effect of the VLFN on the colours and calculate the expected slope of the
branch if the VLFN is indeed motion along the branch. Here the $I_i$'s and
$a_i$'s are the background subtracted count rates and the rms variability
amplitude of the VLFN in the i{\it th} energy band. Given that the {\it
fractional} rms amplitude $r_i \equiv a_i/I_i$, the expected slope 
$\theta_{VLFN}$ of a branch due to VLFN variability is: 

\[ \theta_{VLFN}
= \frac{(I_4 + a_4)/(I_3 + a_3) - I_4/I_3}{(I_2 + a_2)/(I_1 + a_1) - I_2/I_1 }
= \frac{I_4/I_3}{I_2/I_1} \frac{(r_4 - r_3)}{(r_2 - r_1)} 
\frac{(1 + r_1)}{(1 + r_3)},
\]

which should be identical with the measured slope $\theta_{CD}$ 
on the colour-colour diagram: 
\[Z-track\ slope = \theta_{CD} = \frac{\Delta C_H}{\Delta C_S}. \] 
As $r_1$ and $r_3$ are
generally $\ll$1 we can approximate the factor on the right in the  
expression for $\theta_{VLFN}$ with unity
and express the above identity in terms of the relative slopes $\beta_{VLFN}$
and $\beta_{CD}$: 

\[Relative\ variability\ slope = \beta_{VLFN} \cong \frac{r_4 - r_3}{r_2 - r_1}
\equiv \theta_{CD} = 
\frac{\Delta C_H/\overline{C}_H}{\Delta C_S/\overline{C}_S} = \beta_{CD} \]

We are able to test the equality of the Z-track relative slope $\beta_{CD}$
 and the
expected relative slope due to VLFN variability ($\beta_{VLFN}$) for 
two points on the flaring branch; at the mid FB and the
top of the FB (points E and F in Fig\,1).  Only at these points were we able 
to measure
the VLFN rms amplitude in all energy bands. At other places on the Z track
there is not enough signal to measure the VLFN in the lowest energy band. In both
positions on the FB we find that the measurements are consistent (90\%
confidence) with the VLFN being solely due to motion along the Z track. 
However, the
errors are large enough to encompass a factor of 1.5 difference between the
measured Z track relative slope ($\beta_{CD}$) and $\beta_{VLFN}$. 

If the VLFN is solely a product of motion along the Z track then we expect
that within each branch the power spectra of \Sz\ to be a scaled version of
the VLFN component in the power spectra of count rate. This is  because  the
relation between \Sz\ and count rate is approximately linear (Fig.\,3).
Although we cannot, with the present data, measure the power spectrum of \Sz\
on time scales on which the VLFN is typically observed, i.e., $<$200\,sec, we
can measure the count rate power spectra on longer time scales and compare them
to the \Sz\ power spectra. 

Fourier transforms of the \Sz\ values were performed using the fast
implementation of the Lomb-Scargle method for unequally spaced data (Press
\etal\ 1992 and references therein). No data windowing was applied.  Each data
set was taken individually. In each \Sz\ power spectrum an unbroken power law
component was found. The \Sz\ power law index ranged from 1.3 to 2.2 with no 
discernible trend with respect to branch. The average index was
1.7$\pm$\,0.05, similar to that of the VLFN found on shorter (128\,sec)
time scales. 

We have made a direct comparison between the power spectra of count rate and
\Sz\ for two data sets; one on the NB (11-12 March 1986) and one mostly on the
FB (Aug 24 1985). The long (4096\,sec) count rate power spectra (Fig.\,15\,top
panels) show that the power law form of the VLFN component is unbroken out to
time scales of more than 1000\,sec. Working from the other direction, the
power spectra of \Sz\ show an unbroken power law on time scales from about a 9
hours down to a few hundred seconds (Fig.\,15\,bottom panels). The fitted
(VLFN+HFN model) power law slopes of the VLFN component in the count rate power
spectra from the NB and FB are 1.51\,$\pm$\,0.04 and 1.76\,$\pm$\,0.03
respectively over the range 8$\times10^{-4}$ to 0.2\,Hz. 
For the \Sz\ power spectra the power law slopes measured over the $\sim$0.01
to 3$\times10^{-5}$\,Hz range are 1.65\,$\pm$\,0.25 and 1.50\,$\pm$\,0.20 at
on the normal ($\overline{\Sz}$=1.2) and flaring ($\overline{\Sz}$=2.2)
branches. So, the power law indices of the count rate power spectra and the \Sz\
power spectra are consistent with one another on both branches and also
consistent with the $\alpha_{VLFN}$ at the same positions as plotted on
Fig.\,5. Since we have measured similar power indices in two sets of power 
spectra, spanning frequency ranges that overlap from $\sim$0.1 to
$\sim10^{-3}$\,Hz we conclude that it is likely that we are dealing with the
same phenomenon, i.e., motion along the Z track. A similar inference is drawn
from the results of Scargle \etal\ (1993) who have shown, using wavelet
analysis and a scalegram that the VLFN extends, remaining approximately
self-similar and hence being described by a constant slope power law (Scargle
\etal\ 1993), from time scales of about 16\,sec to at least 9 hours. Note that
Scargle \etal\ used an observation (25 Feb 1985) covering both the NB and FB. 

We combined data from consecutive days to produce \Sz\ power spectra extending
to time scales near a day using data taken on 11-13 March 1986 (mostly NB plus
a little HB) and 24-25 Aug 1985 (mostly FB). In the case of the March 1986
data (the longest and most complete data set) we found an unbroken power law
slope with $\alpha$=1.3\,$\pm$0.1, over the range 2$\times10^{-5}$ to
$\sim$\,0.01 Hz., i.e., between time scales of 10\,sec and nearly 14\,hours.
This slope is in the range of the power law indices of the VLFN on the upper
NB/HB ($\alpha$=1.1--1.5, see Fig.\,5.). The 1985 data set is more disjointed
than that of 1986 and so suffers more from windowing. The slope was found to
be 1.4\,$\pm$0.05 over the span 4$\times10^{-5}$ ($\sim$\,7\,hours) to
2$\times10^{-2}$ (50\,sec). This slope is lower than the average slope
(1.7\,$\pm$\,0.05) of the VLFN on the FB. 

Thus the power law shape of the power spectrum of \Sz\  extends unbroken to at
least a 14 hours time-scale with no sign of a turn-over. The power spectrum
must turn-over at some point. The long term behaviour of \sco\ is known to
consist of periods of flaring activity lasting 2-3 days separated by periods
of about 10 days of quieter activity, i.e., the NB/HB. This pattern became
evident from numerous monitoring campaigns; the most recent being with BATSE
(Mc\,Namara \etal\/ 1993) and the RXTE/ASM. Thus the turnover time-scale is
likely to be several days. 

Recently, three alternative models dealing specifically with the VLFN have
been proposed. All three rely on self organized criticality to explain the
power law slope of the VLFN. The classic example of a system that exhibits a
self organized critical (SOC) state is an idealized pile of sand.  As more and
more sand is added, the pile will reach a point when the addition of even a
single grain will set off an avalanche which returns the pile to below the
critical state. SOC models have been applied to a wide range of natural
phenomena. Their main strength is that they naturally produce power-law power
spectra for a wide range of parameters. 

Bildsten (1995) has proposed that the VLFN in Z and atoll sources can be
identified with unsteady nuclear burning on the neutron star's surface. This
is a self organized  critical system (Bildsten, 1995, {\it private
communication}). Two relevant observational consequences of this model are
that the burning luminosity (L$_{B}$) is expected to scale with $\dot{M}$ and
the power law index increases with accretion rate. Thus, the overall behaviour
of the VLFN can be explained by this model. However, there are observations
that it fails to explain.  The power spectrum computed by Bildsten  is not a
power law but has a characteristic timescale in contrast to our observation
the the VLFN power spectrum which is well represented by a power law over a
wide range of timescales. Also, the energy output of nuclear burning is of
course limited. Matter falling onto a neutron star releases about 100 MeV per
proton. Hydrogen burning produces 4\,MeV/proton while helium burning alone
yields 0.9\,MeV/proton and other fusion reactions up to iron can yield a
further 6 MeV/proton. Therefore, being optimistic, up to 11\% of the accretion
luminosity can be released by nuclear burning. The high  (up to
$25\pm_{3}^{4}$\,\%\,rms. See Fig.\,5) VLFN amplitude seen on the upper FB
strains the model. What fractional rms amplitude is measured as VLFN depends
upon the the shape of the light curves from the individual nuclear burning
events and their coherence (Bildsten 1995). However, these properties are
severely constrained by the observed power law slope of the VLFN down to
$\sim$1\,m\,Hz. For example, more spiky wave-forms yield a higher fractional
rms amplitude, but also a steeper power law slope at a higher turn-over
frequency for the VLFN. One possible way out of these constraints is that the
X-ray energy spectra of the nuclear and the accretion flux are different, with
part of the accretion-generated flux below the observable X-ray energy range.
This could also explain why the VLFN rms amplitude spectrum increases with
energy while most of the flux from nuclear burning is expected at low
energies. However, postulating different energy spectra is entirely ad-hoc. 
Another problem with the idea that the VLFN is due to nuclear burning is that
it  dissociates motion through the Z track, and therefore $\dot{M}$, from the
brightness variations comprising the VLFN. This means that on some time-scale
small \Sz-variation related brightness changes must merge ``invisibly" into
the VLFN brightness changes so that the power spectrum does not show a break. 

In a model for the power law component of BHC power spectra which may also be
applicable to the VLFN, Mineshige \etal\ (1994a) sited the variability in the
accretion disk. Their approach was to consider the accretion process through a
disk as a self organized critical (SOC) process. Initial simulations produced
a power law index of 1.8 which is similar to the power law index of the VLFN
of AGN and Z-sources. Further simulations (Mineshige \etal\ 1994b, Takeuchi
\etal\ 1995) including diffusion, i.e., the gradual motion of the material
inward through the disk, reduced the power index to values as low as
$\sim$1.1. The power law slope also depends upon the inclination of the disk
(Abramowicz \& Bao 1994). So far, there is no indication of how the amplitude
of the variations or the power law index depends upon average $\dot{M}$.

The ``dripping handrail" model for the VLFN and QPO, by Scargle \etal\ (1993),
and Young and Scargle (1996) is very similar in concept to that of Mineshige
\etal. Their simulations were able to reproduce not only the VLFN shape but
also can produce QPO whose frequency corresponds to the refill time-scale of
the inner disk. The frequency of the model QPO was proportional to accretion
rate which does not seem to fit the observations. 

{\it 4.2: HFN.} 

We have detected HFN in all branches for \sco, smoothly decreasing in strength
(Xe band: 5--18\,keV) from 7.5\%\,rms in the HB, to 6.3\%\,rms at the NB/FB
vertex, and then to 5.5\%\,rms (Xe band) at the end of the FB. Hertz \etal\
(1992) also found a decrease in HFN strength along the FB; from
5.4\,$\pm$\,0.4\%\,rms (5.8--15.7\,keV) near the NB/FB vertex to
1.6$\pm$0.1\%\,rms in the upper FB. The difference in measured strength can be
attributed mostly to the difference in the energy responses of the EXOSAT ME
and Ginga LAC instruments. Other Z-sources also show a decrease in HFN
strength as the source moves from the HB to the FB (Hasinger \& van der Klis
1989). The EXOSAT instrumental broad band component makes only a  small
contribution because it is compensated by our slight overestimate of the
Poisson noise level in the power spectra (see Sect.\,2.3 for a full
discussion). Because of these instrumental uncertainties one must be cautious
in interpreting HFN amplitudes and frequency extents measured using EXOSAT. 

The HFN was weaker in rms variability by a factor of $\sim$0.5 in the  Ar
energy band (1--15\,keV) than in the Xe band (5--18\,keV). A similar strong
energy dependence was found by Hertz \etal. (1992) where the HFN was, roughly,
twice as strong in the hard (5.8--15.7\,keV) band as in the soft
(1.2--5.8\,keV) band. So far, the energy dependence of the HFN has not been
measured in any other Z-source. 

Because of its weakness and the uncertainties in deadtime correction the HFN
has in other Z-source analyses always been assumed to have $\alpha$$_{HFN}$=0.
We could measure this quantity for \sco. The average power law slope was
slightly positive (0.066$\pm$0.001) which indicates an on average  ``red
noise" shape. There is only slight evidence for a change of slope with \Sz.
The slope is higher (0.071$\pm$0.012) in HB \& NB and (0.02$\pm$0.04) on the
FB. 

We find evidence that the HFN cut-off frequency is affected by the NB/FB
transition. Near this vertex it undergoes a drop from $\sim$75\,Hz to
$\sim$35\,Hz for $\alpha$$_{HFN}$$=0$. This drop occurs within a span of
$<$0.3 in \Sz. We have no information on its timewise behaviour near the NB/FB
vertex. In both the HB-NB and FB the cut-off frequency appears to increase
with \Sz. No other Z source has sufficient count rate with EXOSAT or Ginga to
measure the \Sz\ dependence of the HFN cut-off frequency.

{\it 4.3: LFN.} 

In the HB, the LFN has an amplitude (0.001--1\,Hz) of 2.8\,$\pm$\,0.2\,\%\,rms
and a slope $\alpha=-1.55$$\pm$0.48. Soon after entering the normal branch the
LFN amplitude decreases, becoming undetectable (0.4\%, 2\,$\sigma$ upper
limit) at \Sz=1.4. At the HB/NB vertex the power law index increases rapidly
while the cut-off frequency increases to $>$8\,Hz. This represents a  change
in shape; the LFN becomes broader. 

We find that the LFN is stronger at higher energies. Similar hard spectra for
the LFN have been found from GX~340+0 (Penninx \etal\ 1991) and GX~5-1 (Lewin
\etal\ 1992). 

The shape (peaked or red) of the LFN is one of several characteristics that
seem to split the Z-sources into two groups. \sco\ and GX~17+2 have ``peaked"
LFN while in Cyg~X-2, GX~5-1 and GX~340+0 the LFN is a ``red" noise component
(Hasinger and van der Klis 1989, Penninx \etal\ 1990 \& 1991, and Kuulkers
\etal\ 1994, 1996, 1997). Inclination as well as neutron star magnetic field
strength have been suggested as the underlying difference between the two
groups. Kuulkers \etal\ 1994, 1996 discuss the inclination idea and give
evidence for two groups. Psaltis, Lamb \& Miller, (1995) discuss the magnetic
field hypothesis. The detailed behaviour of the LFN, as a function of \Sz, may
not follow this division of sources. For Cyg~X-2 (high inclination group)
there is a small rise in LFN amplitude just before entering the NB (Fig.\,9 of
Hasinger 1991). However, for GX~5-1, which is of the same group, the LFN
amplitude decreases monotonically (Kuulkers \etal\ 1994 and Lewin \etal\
1992). In GX~340+0 the LFN also amplitude decreases monotonically (Kuulkers
\&\ van der Klis  1996) while for GX~17+2 there are insufficient HFN
measurements (Kuulkers \etal\ 1996) to tell one way or the other. Also for
\sco\ we cannot tell, because of the large errors associated with the fits,
whether the LFN rms amplitude reaches a local maximum, as a function of \Sz,
in the HB. 

{\it 4.4: HBO.} 

No HBO were found in this study of \sco. An upper limit of 3.1\%\,rms (90\%
confidence, Xe band: 5--18\,keV) is set. This limit is set by the strongest
candidate power spectral peak found in the HB data. This peak was near 35\,Hz
with no evidence for an brightness dependence in its frequency. In the cases
of GX~5-1 and Cyg~X-2 the HBO and LFN are  similar in amplitude (rms ratio =
0.75--1.0) at all energies between 1 and 10\,keV. If this holds for \sco, we
expect to see HBO with an amplitude near 2--3\%\,rms (ie comparable with our
measured LFN amplitude). Our upper limit is consistent with this estimate. In
fact, the HBO of GX~17+2, the Z-source most similar to \sco\, has relatively
weak HBO, i.e., the HBO/LFN amplitude ratio is $\sim$0.5 (Hasinger and van der
Klis 1989, confirmed by Kuulkers \etal, 1996). If this ratio holds for \sco\
then we expect the HBO to have an amplitude of only $\sim$1.5\%\,rms. Thus, it
is not surprising that HBO has not been detected from \sco\ during the only
known HB observation.

A HBO-like feature has recently been discovered in the NB of \sco\ (van der
Klis \etal\, 1997) with a frequency near 45\,Hz, FWHM of 15-55\,Hz and an rms
amplitude $\sim$1\%. This feature is in line with the above amplitude limits.

{\it 4.5: NBO/FBO.} 

The rms amplitude of the \sco\ N/FBO increases steadily along the NB-FB till
it is too broad to be  distinguished from the HFN and Poisson noise. GX~17+2,
the only other source where both the NBO and FBO have been observed shows 
the frequency and amplitude higher in the FB than in the NB. The QPO
at the vertex is near 7-8\,Hz (Pennix \etal\ 1990).
For GX~5-1,
there is a strong increase in NBO strength, but no noticeable increase in
frequency with \Sz. (Lewin \etal\ 1992, Kuulkers \etal\ 1994). Cyg~X-2 shows
an increase in NBO frequency from 5.5 to near 7\,Hz in the lower FB but the
strength remains constant or even decreases (Hasinger 1993, {\it private
communication}). For GX~340+0 the NBO amplitude and frequency are constant
within the large (1.5\,\%\,rms and 1\,Hz) scatter. It seems that different
sources behave differently as a function of $\dot{M}$. The fact that these
differences do not correlate with the sub-grouping of Z-sources indicates that
not all timing differences are governed by one single parameter such as
inclination (Kuulkers \etal\ 1994,1996) or magnetic field strength (Psaltis,
Lamb \& Miller 1995). 

It has been suggested that the NBO are the result of a radiation-force/opacity
feed-back loop within a spherical flow region (Lamb 1989). Alternatives are
sound waves in a thick disk (Alpar \etal\ 1992) or g-mode oscillations in the
neutron star (Bildsten \& Cutler 1995, Bildsten \etal\ 1996). 

There are concerns about how to excite a single g-mode oscillation at the
neutron star's surface with sufficient amplitude to be detectable
\footnote{Bildsten \etal\ 1996 show that g-modes can only be sustained in
slowly rotating neutron stars which seems unlikely given the identification of
$\approx$360\,Hz burst QPO's with neutron star spin frequencies.}. The main
observable difference between g-mode oscillations and other models is that
when a second harmonic is excited it should have a frequency a factor
$\sqrt{3}$ rather than a factor of two greater than the fundamental. The
frequency change across the NB/FB vertex provides a potential discriminator.
However, our data is consistent with either model.

We will now concentrate on the  interpretation of our observations in terms of
the radiation-force feedback model. The spherical flow region (Lamb 1989) is
expected to become  prominent only above $\sim$0.8$L_{Edd}$.  At luminosities
$\geq L_{Edd}$ the smooth flow will be broken up by ``bubbles" of photons
being pushed out by buoyancy. At some point any oscillation modes will be
destroyed. The NB/FB vertex is thought to occur within a few percent of the
Eddington limit. 

This scenario nicely explains the limited extent over which the QPO are
visible. We find QPO from \sco\ over only 17\% of the Z track in the lowest
third of the NB and the lowest 10\% of the FB measured in terms of \Sz. The
\Sz\ positions at which the QPO first become measurable are practically the
same between sources; i.e., for \sco: \Sz=1.5, GX~5-1: \Sz=1.5 (Kuulkers
\etal\ 1994), Cyg~X-2: rank=1.6 or \Sz=1.3 (Hasinger, 1993, {\it private
communication}, GX~17+2: \Sz=1.4--1.5 (Kuulkers \etal\ 1997) and for GX~340+0:
\Sz=1.2 (Kuulkers \& van der Klis 1996). The FBO of GX~17+2 are limited to
about the lowest 15\% of the FB (Penninx \etal\ 1990). 

At the high luminosities observed from Z-sources radiation forces increase the
dynamical time-scale of the inward material flow.   The fastest time-scale
seen for the 7\,Hz to $>$12\,Hz transition was $\leq$\,90\,sec. Although this
is the fastest time-scale reported for the change in NB/FBO frequency it is
still much longer than the estimated dynamical time-scale of $\approx$1\,sec
(Lamb 1989), and so cannot constrain models. 

The most remarkable result presented here is the rapid change with respect to
both \Sz\ and time in NB-FB QPO frequency at the NB/FB vertex. The
frequency-position diagram shows that the QPO frequency is very sensitive to
position near the vertex. 

This sensitivity to Z-track position has not been previously noted. The
frequency-intensity diagram used by Priedhorsky \etal\ (1986) emphasizes the
smooth continuous nature of the transition but not its restricted range upon
the Z track. With hindsight, Fig.\,7 of van der Klis \etal\ (1987) also shows
a jump in QPO frequency near the NB/FB vertex from near 7\,Hz to 12\,Hz. Here
the QPO frequency is plotted against the black body luminosity ($L_{BB}$) as
determined from two component fits to the X-ray energy spectrum. Both $L_{BB}$
and \Sz\ are postulated to be directly related to the mass accretion rate onto
the neutron star. The sense of the change in QPO frequency with $L_{BB}$ is
the same as with \Sz. The range in $L_{BB}$ over which the QPO frequency
changed from 6.5 to 12\,Hz occurred is only $\sim$~13\% of the range in which
QPO were measurable. This is similar to the $\leq$8\% range in \Sz. 

The jump in N/FBO frequency at the NB/FB vertex can be viewed as the result of
either a sensitive dependence of frequency upon the underlying controlling
parameter, presumably $\dot{M}$ (but see the next paragraph), with \Sz\
approximately proportional to this controlling parameter.  An alternative view
is that the relationship between \Sz\ and $\dot{M}$ is different near the
NB/FB vertex, with a proportionally larger change in $\dot{M}$ as \sco\ moves
the short distance around the vertex as compared to motion in other places on
the Z track. Either way, the NB/FB vertex represents a critically important
Z-track position. 

Within the context of a unified model of the flow about low mass neutron stars
(Lamb 1989), modeling of the Z track (Psaltis, Lamb \& Miller, 1995) suggests
that $\dot{M}$ changes smoothly along the Z track. Thus in this model the
sensitivity of QPO frequency on \Sz\ is not due to changes in the relation
between \Sz\ and $\dot{M}$. The parameter controlling the behaviour of the
flow onto the neutron star is the difference between the accretion rate and
the Eddington critical rate. The radial flow region becomes prominent only
when the luminosity is within 20\% of L$_{Edd}$. Much closer (within 2\%) to
the Eddington accretion rate the flow conditions are expected to become more
complex. The initial model for the N/FB QPO of Fortner, Lamb and Miller (1989)
considered only radial ($\ell$=1 n=1) oscillations in a smooth spherical
inflow region at luminosities not too close to L$_{Edd}$. This model
successfully accounts for the observed low frequency of the QPO and in the
case of Cyg\,X-2 the dependence of the NBO amplitude and relative phasing with
energy (Miller \& Lamb, 1992 and references therein).  If the outer radius of
the cool flow is constant the model predicts that the NB/FBO frequency
decreases with increasing mass flux in contrast to our observations. However
the behaviour of the outer radius is difficult to model and hence uncertain. 
Recently Miller \& Park (1995) have considered non-radial oscillations within
the inflow region. At accretion rates$\sim$0.8 to 0.98 L$_{Edd}$ any
non-radial modes are well damped. But as the luminosity increases to within
2\% of L$_{Edd}$ they find that the higher frequency non-radial modes (e.g.
$\ell$=1, n$>$1 or $\ell$=2, n$>$1) are much less damped and can dominate the
lower order oscillations and so the frequency rises.  The span in luminosity
over which the higher frequency modes begin to dominate is very similar to the
range in \Sz\ over which the NB/FBO oscillations increase in frequency. These
higher order modes continue to be excited as the source exceeds the Eddington
limit but will be suppressed as the flow becomes fragmented. In this picture,
the anomalous frequency excursions we observe may be the result of momentary
switches between oscillation modes. 
 
{\it 4.6: Is $\dot{M}$ all there is\,?} 

Although direct searches for hysteresis failed we have found two  indications
that position along the Z track is not the only parameter  governing the
behaviour of \sco.  First the minimum values of count rate reached during NB/FB
transitions are variable on time scales of several thousand seconds. This
behaviour would  contribute to the increased scatter of intensities and QPO
frequency seen in  the lowest parts of the NB and FB as seen on a
frequency-intensity diagram. Secondly; the two anomalous frequency excursions
that clearly break the paradigm of a one-to-one relation between spectral
state as determined from position in a colour-colour diagram and timing
properties.

The common features of these cases are that they occur at or near the NB/FB
vertex (i.e. L$_{Edd}$) and that their time scales are short. There thus may be a link
with the changes in the flow expected near L$_{Edd}$. The exact nature of the
flow and the resulting QPO could depend upon other factors in addition to
$\dot{M}$. For example, which of the QPO modes dominates may depend upon the
optical depth of the flow as \sco\ reaches L$_{Edd}$. Observationally, the way
to proceed is to examine the X-ray spectral, QPO and noise properties on
shorter time scales than was possible with our data, especially near the NB/FB
transition where we have shown that the QPO properties are particularly
sensitive to the underlying controlling parameters. 

{\large\bf Acknowledgments.} 

We gratefully acknowledge many useful discussions with Erik Kuulkers, Brian
Vaughan, Lars Bildsten, Fred Lamb and Guy Miller. This work was supported by
the Netherlands Organization for Scientific Research (NWO) under grant PGS
78-277 and S.D. was also partially supported by grant ERB-CHRX-CT93-0329 of
the European Commission (HCM program).\\

{\large\bf References.}

{\small

\rf{Alpar M.A., \& Shaham J., (1985) {\it Nature}, {\bf 316}, 239.}

\rf{Alpar M.A., Hasinger G., Shaham J., Yancopoulos S., (1992)
Astron.~Astrophys., {\bf 257}, 627.}

\rf{Ambramowicz M., \& Bao G., (1994) Publ. Astron. Soc. Japan {\bf
46}, 523.}

\rf{Augusteijn T., Karatasos K., Papadakis M., Paterakis G., Kikuchi S., Brosch
N., Leibowitz E., Hertz P., Mitsuda K., Dotani T., Lewin W.H.G., van der Klis
M. \& van~Paradijs~J., (1992) Astron. Astrophys., {\bf 265}, 177.} 

\rf{Andrews D., \& Stella L., (1985) EXOSAT Express, {\bf 10}, 35.} 

\rf{Berger M., \& van der Klis M., (1994) Astron. \& Astrophys., {\bf 292},
175.} 

\rf{Berger M., \& van der Klis M., (1998) Astron. \& Astrophys., Astron. \& Astrophys., {bf 340}, 143.} 

\rf{Bildsten L., (1995), Astrophys. J., {\bf 438}, 852.} 

\rf{Bildsten L., \& Cutler C., (1995) Astrophys. J., {\bf 449}, 800.} 

\rf{Bildsten L., Ushomirpsky G., \& Cutler C., (1996), Astrophys. J., {\bf 460}, 827.}

\rf{Deeter J.E., (1984) Astrophys. J., {\bf 281}, 482.} 

\rf{Fortner B., Lamb F.K., \& Miller G.S., (1989) {\it Nature}, {\bf 342},
775.} 

\rf{Giacconi R., Gursky H., Paolini F., \& Rossi B., (1962) Phys. Rev. Lett.,
{\bf 9}, 439.} 

\rf{Haberl F., (1992) {\it Legacy}, {\bf 1}, 53.} 

\rf{Hasinger G., (1987a) Astron. Astrophys., {\bf 186}, 153.} 

\rf{Hasinger G., (1987b) in {\it The Origin and Evolution of Neutron Stars},
D.J. Helfand \& J.-H. Haung (eds.) IAU Symp. 125, p333.} 

\rf{Hasinger G., (1991) in {\em Particle Acceleration near Accreting Compact
Objects},\  J. van Paradijs, M. van der Klis, \& A. Achterberg (eds.) \  Royal
Netherlands Academy of Arts and Sciences., North Holland, Amsterdam, p23.} 

\rf{Hasinger G., Langmeier A., Sztajno M., Tr\"umper J., Lewin W.H.G., \& White
N.E., (1986), {\it Nature}, {\bf 319}, 469.} 

\rf{Hasinger G., Priedhorsky W.C., \& Middleditch J., (1989)
Astrophys. J., {\bf 337}, 843. (HPM89).} 

\rf{Hasinger G. \& van der Klis M., (1989) Astron. Astrophys., {\bf 225}, 79.
(HK89).} 

\rf{Hasinger G., van der Klis M., Ebisawa K., Dotani T. \& Mitsuda K., (1990)\ 
Astron. Astrophys., {\bf 235}, 131.} 

\rf{Hertz P., Vaughan B., Wood K.S., Norris J.P., Mitsuda K., Michelson P.F.,\ 
\& Dotani T., (1992) Astrophys.~J., {\bf 396}, 201.} 

\rf{Hjellming R.M., Stewart R.T., White G.L., Strom R., Lewin W.H.G., Hertz P.,
Wood K.S., Norris J.P., Mitsuda K., Penninx W., \& Paradijs J., (1990)
Astrophys.~J., {\bf 365}, 681.} 

\rf{Jain A., Hasinger G., Pietsch W., Proctor R., Reppin C., Tr\"umper J.,
Voges W., Kendziorra E.,\  \& Staubert R., (1984) Astron. Astrophys., 
{\bf 140}, 179.} 

\rf{Kuulkers E., van der Klis M., Oosterbroek T., Asai K., Dotani T.,
van~Paradijs J.,\  \& Lewin W.H.G., (1994), Astron. Astrophys., {\bf 289},
795. Erratum (1995) Astron. Astrophys., {\bf 295}, 842.} 

\rf{Kuulkers E., (1995) Ph.D. Thesis, {\it EXOSAT observations of Z Sources},
University of Amsterdam.} 

\rf{Kuulkers E., van der Klis M., (1995) Astron. Astrophys., {\bf 303}, 801.}

\rf{Kuulkers E., van der Klis M., \& van Paradijs J., (1995) Astrophys.~J.,
{\bf 450}, 748.} 
 
\rf{Kuulkers E., \& van der Klis M, (1996) Astron. Astrophys., {\bf 314}, 567.} 

\rf{Kuulkers E., van der Klis M., \& Vaughan B.A., (1996)
Astron. Astrophys. {\bf 311}, 197.} 

\rf{Kuulkers E., van der Klis M., Oosterbroek T., van~Paradijs J.,\  \& Lewin
W.H.G., (1997), Mon. Not. Roy. Astron. Soc., {\bf 287}, 495.} 

\rf{Lamb F.K. (1989) in Proc. 23rd ESLAB Symp. in {\it Two Topics in X-ray
Astronomy}, J. Hunt \& B. Battrick (eds.), ESA SP-296, {\bf Vol.\,1} p215.} 

\rf{Lamb F.K., Shibazaki N., Alpar M.A., \& Shaham J., (1985) {\it Nature}, {\bf
317}, 681.} 

\rf{La\,Sala J., \& Thorstensen J.R., (1985) Astron. J., {\bf 90}, 2077.} 

\rf{Langmeier A., Hasinger G., \& Tr\"umper J., (1990) Astron. Astrophys., {\bf
228}, 89.} 

\rf{Leahy D.A., Darbro W., Elsner R.F., Weisskopf M.C., Sutherland
F.G., Kahn~S., \& Grindlay J.E.,\  (1983) Astrophys. J., {\bf 266}, 160.} 

\rf{Lewin W.H.G., van Paradijs J., \& van der Klis M., (1988) Spa. Sci. Rev.,
{\bf 46}, 273.} 

\rf{Lewin W.H.G., Lubin L.M., Tan J., van der Klis M., van Paradijs J., Penninx
W., Dotani T., \& Mitsuda K., (1992) Mon. Not. Roy. Astron. Soc., {\bf 256},
545.} 

\rf{Mc\,Namara B., Fitzgibbons G., Fishman G.J., Meegan C.A., Wilson R.B.,\ 
Harmon~B.A., Paciesas W.S., Rubin B.C., \& Finger M.H., (1994).\  {\em Compton
Gamma-ray Observatory}, St Louis 1993,\  M. Friedlander, N. Gehrels \& D.J.
Macomb (eds.), AIP Press, New York, p391.} 

\rf{Middleditch J., \& Priedhorsky W.C., (1986) Astrophys. J., {\bf 306}, 230.}

\rf{Miller G.S., \& Lamb F.K., (1992) Astrophys. J., {\bf 388}, 541.} 

\rf{Miller G.S., \& Park M-G., (1995) Astrophys. J., {\bf 440}, 771.} 

\rf{Mineshige S., Ouchi B., \& Nishimori H., (1994a) Publ. Astron. Soc. Japan,
{\bf 46}, 97.} 

\rf{Mineshige S., Takeuchi M., \& Nishimori H., (1994b) Astrophys. J., {\bf
435}, L125.} 

\rf{Mook D.E., Messina R.J., Hiltner W.A., Belian R., Conner J., Evans W.D,
Strong~I.,\  Blanco V.M., Hesser~J.E., Kunkel W.E., Lasker B.M., Golson J.C.,
Pel~J.,\  Stokes N.R., Osawa K., Ichimura~K., \& Tomita~K., (1975)
Astrophys.~J., {\bf 197}, 425.} 

\rf{Penninx W., Lewin W.H.G., Mitsuda K., van der Klis M., van Paradijs J., \&
Zijlstra A.A.,\  (1990) Mon. Not. Roy. Astron. Soc., {\bf 243}, 114.} 

\rf{Penninx W., Lewin W.H.G., Tan J., Mitsuda K., van der Klis M., \&
van~Paradijs J.,\  (1991) Mon. Not. Roy. Astron. Soc., {\bf 249}, 113.} 

\rf{Petro L.D., Bradt H.V., Kelley R.L., Horne K., \& Gomer R., (1981)
Astrophys. J., {\bf 251}, L7.} 

\rf{Pollock A.M.T., Carswell R.F., \& Ponman T.J., (1986)  poster presented at
the workshop on {\em The Physics of Compact Objects} held in Tenerife, Spain.} 

\rf{Ponman T.J., Cooke B.A., \& Stella L., (1988) Mon. Not. Roy. Astron. Soc.,
{\bf 231}, 999.} 

\rf{Press W.H., Teukolsky S.A., Vetterling W.T., \& Flannery B.P., (1992)\  in
{\it Numerical Recipes: The Art of Scientific Computing} 2nd Edition.\ 
Cambridge University Press. Cambridge.} 

\rf{Priedhorsky W., Hasinger G., Lewin W.H.G., Middleditch J., Parmar A.,
Stella L., \& White N.,\  (1986) Astrophys. J., {\bf 306}, L91.} 

\rf{Prins S., \& van der Klis M., (1997) Astron. Astrophys., {\bf 319}, 498.} 

\rf{Psaltis D., Lamb F.K., \& Miller G.S., (1995) Astrophys. J., {\bf 545}, L137.} 

\rf{Scargle J.D., Steiman-Cameron T., Young K., Donoho D.L., Crutchfield J.P.,
\& Imamura J., (1993) Astrophys. J., {\bf 411}, L91.} 

\rf{Sztajno M., van Paradijs J., \& van der Klis M., Langmeier A., Tr\"umper J.,
\& Pietsch W.,  (1986) Mon. Not. Roy. Astron. Soc., {\bf 222}, 499.} 

\rf{Takeuchi M., Mineshige S., \& Negoro H., (1995)
Publ. Astron. Soc. Japan. {\bf 47}, 617.}

\rf{Turner M.J.L., Smith A., \& Zimmermann H.U., (1981) Spa. Sci. Rev., {\bf
30}, 513.} 

\rf{Ubertini P., Bazzano A., Cocchi M., La\,Padula C., \& Sood R.K., (1992)
Astrophys.~J., {\bf 386}, 710.} 

\rf{van der Klis M., Jansen F., van Paradijs J., Lewin W.H.G., van den Heuvel
E.P.J.,\  Tr\"umper J.E. \& Sztajno~M., (1985) {\it Nature}, {\bf 316}, 225.} 

\rf{van der Klis M., Stella L., White N., Jansen F., \& Parmar A.N., (1987)
Astrophys. J., {\bf 316}, 411.} 

\rf{van der Klis M., (1989a) Ann. Rev. Astron. Astrophys., {\bf 27}, 517.} 

\rf{van der Klis M., (1989b) in {\it Timing Neutron Stars}, H. \"Ogelman \&
E.P.J. van den Heuvel (eds).,\  Kluwer, Dordrecht, NATO ASI Series C {\bf 262},
p27.} 

\rf{van der Klis M., (1991) in {\it Neutron Stars: Theory and Observation},
J.~Ventura \& D.~Pines (eds.),\  Kluwer, Dordrecht, NATO ASI Series C
{\bf344}, p319.} 

\rf{van der Klis M., (1995a) in {\it Lives of Neutron Stars} Alpar M.A,
Kizilo\u glu \"U., \& van~Paradijs~J.,eds.\, Kluwer, Dordrecht, NATO ASI Series
C {\bf 450}, p301.} 

\rf{van der Klis M., (1995b) in {\it X-ray Binaries}, W.H.G Lewin, J.
van~Paradijs  \& E.P.J. van den Heuvel (eds.) Cambridge, U.K., Astrophysics
Series {\bf 26}, p252.} 

\rf{van der Klis M., (1998) in {\it The Many Faces of Neutron Stars}, R. Buccheri,
 J. van Paradijs, M.A. Alpar, Kluwer, Dordrecht, NATO ASI Series.
(astro-ph/9710016). 

\rf{van der Klis M., Wijnands R.A.D., Horne K., \& Chen W., (1997) Astrophys. J., {\bf 481}, L97.}

\rf{Vaughan B.A., van der Klis M., Wood K.S., Norris J.P., Hertz P., Michelson
P.F., van Paradijs J.,\  Lewin W.H.G., Mitsuda K., \& Penninx W. (1994)
Astrophys. J. {\bf 435}, 362.} 

\rf{Vrtilek S.D., Raymond J.C., Garcia M.R., Verbunt F., Hasinger G., \&
K\"urster M.,\  (1990) Astron. Astrophys., {\bf 235}, 162.} 

\rf{Vrtilek S.D., Penninx W., Raymond J.C., Verbunt F., Hertz P., Wood K.,
Lewin W.H.G., \& Mitsuda K., (1991) Astrophys.~J., {\bf 376}, 278.} 

\rf{Walpole R.E., \& Myers R.H., (1978) in {\it Probability and Statistics for
Engineers and Scientists}\  (2nd ed.) Macmillain Pub. Co., New York.} 

\rf{White N.E., Peacock A., (1988) in {\it X-ray Astronomy with EXOSAT},
R.~Pallavicini, \& N.E. White (eds.), Mem. S. A. It. {\bf 59}, 7.} 

\rf{Young K., Scargle J.D., (1996) Astrophys.~J., {\bf 468}, 617.} 

}\newpage

\begin{table}[h]
\caption[]{Observation Log}
\smallskip
\begin{tabular}{@{}|lc|cc|r|cc|c|c|@{}}
\hline
\multicolumn{2}{|c|}{Date and Day Number} & Start  &  End  & Dur.    &\multicolumn{2}{|c|}{Binning~(ms)}    & Branch& Ref\\
\multicolumn{2}{|c|}{     } &\multicolumn{2}{|c|}{Time (UT)} & (sec)   & Time$^{x}$ & Energy$^{y}$ &       & \\
\hline
\multicolumn{9}{|c|}{Observations examined in detail.}\\
 29/30 Aug 1983 & 241/2 & 23:09 & 18:47 & 70\/720 &  8       &2500$^a$   & ~FB \& NB$^h$ & 1 \\   
 03    Aug 1984 & 216   & 06:38 & 12:03 & 19\/328 &  ~8$^c$  &312.5$^b$  & NB$\rightarrow$FB & 2 \\
 25    Feb 1985 &  56   & 03:58 & 13:18 & 32\/104 &  2       & ---   & NB$\rightarrow$FB & 3 \\
 24    Aug 1985 & 236   & 09:02 & 14:50 &  5\/784 &  4       &  4    &  FB    & 4 \\
 25    Aug 1985 & 237   & 13:05 & 20:18 & 25\/976 &  4       &  4    & NB$\rightarrow$FB & 4 \\
 11/12 Mar 1986 & 70/71 & 16:37 & 11:05 & 66\/536 &  1       &  8    & NB$\rightarrow$HB & 5 \\
 13    Mar 1986 &  72   & 02:18 & 12:03 & 35\/160 &  1       &  8    &  mid NB           & 5 \\
~~~~~~~~"       &  72   & 16:13 & 19:03 & 10\/200 & ~\,8$^d$ &  4    &  low NB    & 5 \\
~~~~~~~~"       & 72-73 & 19:09 & 01:12 & 21\/716 &  1       &  8    &  low NB    & 5 \\
\multicolumn{9}{|c|}{Other data searched for frequency excursions.$^g$}\\
 11 Mar 1984    &  71   & 02:06 & 07:37 & 17224 &  8+125$^e$ & 5000$^a$ & FB     & 1\\
 13 Mar 1984    &  73   & 00:59 & 04:36 & 12800 &  8+125$^e$ & 5000$^a$ & mid NB & 1 \\
 26 Aug 1985    & 238   & 10:00 & 14:34 & 15696 &  4         &  4$^f$   & NB     & 4\\
 27 Aug 1985    & 239   & 04:19 & 14:16 & 34328 &  4         &  4$^f$   & FB$\rightarrow$NB & 4\\
\hline
\noalign{\smallskip}
\multicolumn{9}{@{}p{140mm}@{}}{ \rf{x Unless otherwise noted all high time
resolution data is taken with the HTR3 or HTR5 OBC modes, using solely the Xe
filled chambers.  Only the fastest available sampling is listed.}}\\
\multicolumn{9}{@{}p{140mm}@{}}{ \rf{y Unless otherwise stated this column refers
to HER7 data with energy boundaries at 0.9, 3.1, 4.9, 6.6 and 19.5\,keV. Only
the fastest available sampling is listed.}}\\
\multicolumn{9}{@{}p{140mm}@{}}{ $^a$ Multichannel data; affected by overflows.}\\
\multicolumn{9}{@{}p{140mm}@{}}{ $^b$ Multichannel HER5 data.}\\
\multicolumn{9}{@{}p{140mm}@{}}{ $^c$ Ar+Xe combined count rate.}\\
\multicolumn{9}{@{}p{140mm}@{}}{ $^d$ Various combinations of the HTR5 and HER7 modes were used.}\\
\multicolumn{9}{@{}p{140mm}@{}}{\rf{$^e$ Four sets of single channel data (one from each
pair of Xe detectors) at 125\,ms time resolution in addition to the count rate
sum every 8\,ms from all Xe detectors. The corresponding energy resolved data (5000\,ms) is
from the Ar and Xe detectors of each half array separately.}}\\
\multicolumn{9}{@{}p{140mm}@{}}{ $^f$ Four energy channel data (HER7) but with 2 Ar channels and 2 Xe channels.}\\
\multicolumn{9}{@{}p{140mm}@{}}{\rf{$^g$ Observations on 7/8 Aug 1983 where also examined,
but no QPO were found. However, the time resolution and continuity were generally inadequate to detect QPO.}}\\
\multicolumn{9}{@{}p{140mm}@{}}{ $^h$ The FB$\rightarrow$NB transition occurs during a gap in the data.}\\
\multicolumn{9}{@{}p{140mm}@{}}{ {\bf [1]} van~der~Klis~et~al.~(1987)~~~~~~~~~~~~~~~~~{\bf [2]} Pollock \etal\ (1986) }\\
\multicolumn{9}{@{}p{140mm}@{}}{ {\bf [3]} Middleditch~\&~Priedhorsky~(1986)~~~~~{\bf [4]} Priedhorsky \etal\ (1986) }\\
\multicolumn{9}{@{}p{140mm}@{}}{ {\bf [5]} Hasinger, Priedhorsky \& Middleditch (1989)}
\label{OLog}
\end{tabular}
\end{table}

\newpage

\null\hoffset=-2cm\null
\begin{table}[h]
\caption[]{Energy dependence of power spectral components}
\smallskip
\small
\begin{tabular}{|l|c|ll|ll|cl|ll|}
\hline
    & {\normalsize Sz}   &\multicolumn{2}{c}{VLFN} &\multicolumn{2}{c}{{\normalsize LFN}}
           &\multicolumn{2}{c}{{\normalsize QPO}}  &\multicolumn{2}{c|}{{\normalsize HFN}} \\
    &      &\multicolumn{2}{c}{ {\footnotesize \%rms (0.001--1Hz)}}
           &\multicolumn{2}{c}{{\footnotesize \%rms (0.001--100Hz)}} 
           &\multicolumn{2}{c}{ }
           &\multicolumn{2}{c|}{{\footnotesize \%rms (1--500Hz)}}\\ 
    &      &    Ar   &     Xe    &    Ar    &   Xe    &    Ar    &    Xe   &    Ar   &     Xe  \\
\hline
    &      &        &           &           &         &          &         &         &         \\
A & 0.645$\pm$0.075 & $0.9 \pm 0.1      $ & $~1.0 \pm 0.1      $
                    & $1.9 \pm 0.2      $ & $ 3.7 \pm 0.25     $
                    &                       & 
                    & $3.0 \pm 0.45     $ & $ 6.8 \pm 0.3      $\\
    &      &        &           &           &         &          &         &         &         \\

B & 1.473$\pm$0.075 & $0.46 \pm 0.05    $ & $~1.6^{+0.13}_{-0.09}$
                    &                       & $ 1.8^{+1.3}_{-0.2}$
                    &                       & 
                    & $4.0 \pm 0.1      $ & $ 7.0 \pm 0.1      $\\
    &      &        &           &           &         &          &         &         &         \\

C & 1.735$\pm$0.040 & $0.8 \pm 0.4      $ & $~2.5 \pm 0.07     $
                    &                       & 
                    & $1.3 \pm 0.08     $ & $ 3.3 \pm 0.15     $
                    & $4.2 \pm 0.1      $ & $ 6.4 \pm 0.2      $\\
    &      &        &           &           &         &          &         &         &         \\

D & 2.161$\pm$0.072 & $1.8 \pm 0.1      $ & $~3.1^{+0.3}_{-0.2} $
                    &                       & 
                    & $<$2.12 (1$\sigma$)   & $ 6.5^{+0.4}_{-0.2} $
                    & $2.57 \pm 0.65    $ & $ 5.55 \pm 0.35   $\\
    &      &        &           &           &         &          &         &         &         \\

E & 2.753$\pm$0.151 & $3.9 \pm 0.2      $ & $~6.9^{+0.9}_{-0.7} $
                    & $1.15 \pm 0.7     $ & $ 3.0^{+0.7}_{-0.3} $ 
                    &                       & 
                    & $2.41 \pm 0.75    $ & $ 5.5^{+0.3}_{-0.4} $\\
    &      &        &           &           &         &          &         &         &         \\

F & 4.108$\pm$0.301 & $9.3^{+2.7}_{-1.7}  $ & $23.0^{+4.4}_{-2.9}  $
                    &                       & 
                    &                       & 
                    & $4.1 \pm 0.7      $ & $ 5.25^{+1.4}_{-0.5} $\\
    &      &        &           &           &         &          &         &         &         \\
\hline
\noalign{\smallskip}
\multicolumn{10}{@{}p{175mm}@{}}{ The amplitudes for the Ar data were found by
fixing all other parameters to the values from the Xe fits. Full fits to the Ar
and Xe power spectra showed that all non-amplitude parameters were the same
within $<$2{$\sigma$}. However, the errors were much larger for the Ar power
spectra.}\\
\multicolumn{10}{@{}p{175mm}@{}}{ }\\
\multicolumn{10}{@{}p{175mm}@{}}{ The "LFN`` of spectrum E is an extra QPO/LFN like component at $\simeq$5.5\,Hz.}
\label{OLog}
\end{tabular}
\end{table}

\newpage

\hoffset=0pt

{\bf Fig.\,1.} The colour-colour diagram of \sco\ showing the complete
Z-track. The data were gathered using the HER7 mode, which gives data in 4
energy channels. The integration time is 196 sec per point. Typical errors are
shown at three different places in the Z-track. The horizontal, normal and
flaring branches are labeled as HB, NB and FB, respectively. The solid curve
is the Z-track representation used to calculate \Sz\ and \Dz\ (see text). It
is marked off at intervals of 0.5 in the values of \Sz. The frequency of the
QPO is given at 4 places along the Z-track. Higher on the Z-track than the
points marked ``6\,Hz", and ``20\,Hz" the QPO are not detectable. Also
labeled, as A,B,C,D,E,F,  are the positions at which the 6 representative
power spectra of Fig.\,4 were taken. 

{\bf Fig.\,2.} The colour-colour diagram based upon the HER5 mode data taken
on 3 Aug 1984. The colour-colour points were calculated in the same energy
bands as in Fig.\,1.  The solid curve is the Z-track representation of the 3
Aug 1984 data. The dash-dotted curve is the Z-track from Fig.\,1. The
differences in Z-track position and orientation are within the uncertainties
in the colours caused by uncertainties in the background and changes in the
gain of the Ar counters. 

{\bf Fig.\,3.} The Xe count rate (background subtracted and scaled to 8
detectors) versus Z-track position (\Sz). The individual data points
correspond to intervals in time or \Sz\ where power spectra were calculated.
Near the NB/FB vertex (\Sz=\,2) linear fits were made between count rate and
position. The slope on the NB side of the vertex is $-$2120$\pm$70 cts/\Sz-unit
and on the FB side 1500$\pm$200 cts/\Sz-unit.  The count rate at the vertex is
2239$\pm$26 cts/sec/8~detectors. 

{\bf Fig.\,4.} Representative power spectra. The expected Poisson level has
been subtracted and corrections have been made for the effects of differential
deadtime. The errors are 1 sigma. The positions on the Z-track where these
spectra were taken are shown on Fig.\,1. At each position two spectra are
shown; one from the 2--20\,keV Ar data (open squares, lighter smooth curve)
and the other from the 5--35\,keV Xe data (filled circles, bolder smooth
curve). The smooth curves represent the best fits to the power spectra.  In
Fig.\,4C, two fits are given for the Ar data; one with the HFN cut-off
frequency fixed at that derived from the Xe data and the other with the
cut-off frequency as a free parameter. In Fig.\,4E, there are also two fits
for the Ar power spectrum, i.e., with and without an extra peaked noise
component that is required to fit the Xe data.

{\bf Fig.\,5.} The parameters of all broad band noise components as a function
of Z-track position (\Sz). The HB/NB vertex at \Sz=\,1 and the NB/FB vertex at
\Sz=\,2 are marked by dotted vertical lines. Each point was derived from a fit
to an average of a number of power spectra obtained at the \Sz\ value shown.
Each individual power spectrum was calculated from 128\,sec of Xe band
(5--50\,keV) timing data. The fit always included VLFN and HFN components and,
if required, either a LFN or a QPO component. The error bars in \Sz\ show the
span covered during each fit. The vertical error bars  represent the
1\,$\sigma$ single-parameter errors as determined from a scan in $\chi^2$
space. The rms amplitudes were corrected for differential deadtime effects. 

{\bf Fig.\,6.} The relative width of the QPO (FWHM/QPO frequency) as a
function of \Sz. The QPO are most coherent near \Sz=\,1.75. 

{\bf Fig.\,7.} The variation of QPO properties with Z-track position (\Sz).
The same symbols as in Fig.\,1 are used to identify data from different
observations. Panels {\bf a} and {\bf b} show the fractional rms amplitudes
(corrected for differential deadtime and frequency binning) for the Ar+Xe
(HER5; 3~Aug 1984) and Xe only (HTR; 1985--86) data respectively. Note that
the QPO is always weaker at lower energies. Panel {\bf c} shows the QPO
frequency. Here the Xe-only and the Ar+Xe data are shown together. The
transition between the NB and FB is not resolved. It occurs within $\pm$0.05
\Sz-units of the NB/FB vertex. Panel {\bf d} shows the frequency-intensity
diagram. The count rates are for the Xe data with background subtraction and
scaled to the full array of 8 detectors.  In this panel there are extra points
not present in the other panels. These points (represented by error bars only)
are from 25 Feb 1985, when there was excellent timing, but no energy-resolved
data. Two data points are circled in each panel. These points correspond to
the 13 Mar 1986 anomalous excursion to higher QPO frequencies (see text). 

{\bf Fig.\,8.}  The detailed behaviour of \sco\ during the anomalous frequency
excursion of 13 Mar 1986. Time is measured from 16:13 UT when the observation
began. The time of highest QPO frequency and X-ray count rate is labeled
``Peak", the two count rate dips associated with 8\,Hz QPO are labeled ``Dip".
The top two panels show the measurements of the HER7 colours as used in the
colour-colour diagram of Fig.\,1 and used to calculate the values of \Sz. The
thick line represents the average colours over the same time intervals as used
to measure the power spectra. This is the same binning as used in Figs.\,7, 9
and 10.  The thin line shows the colours as measured with a 32\,sec
integration time.  The 3rd panel shows the values of \Sz\ as calculated from
these colours. The \Sz\ values represented by the thick line are those
calculated from the colours measured with a 16\,sec integration time, averaged
over the selected data sections and processed using the correction/replacement
algorithm described in Sect.\,2.1. The thin line shows a 3$\times$32\,sec
point binning of the \Sz\ values before any corrections were applied.
Comparison of the two curves shows that the low \Sz\ measurements at the time
of the frequency excursion (1.5$10^4$ sec) do not result from our
correction/replacement method. The 4th panel shows the Ar/Xe hardness ratio
(colour). Whereas the individual HER7 colours do not show any clear changes
this hardness ratio clearly show dips at the times of the count rate dips. The
5th panel shows the variations in QPO frequency. The line represents
measurements made by fitting the power spectra. One sigma errors are
indicated. The points result from a completely different method for measuring
the QPO frequency, where a threshold power level was set and the mode
frequency of the powers above this threshold was found. The formal errors are
of the order of 3--4\,Hz when a 32\,sec integration time is used. The bottom
panel shows the Xe and Ar count rates as defined in Sect.\,2. 

{\bf Fig.\,9} The frequency-intensity diagram for \sco, including the  track
taken during the frequency excursion on 13 Mar 1986. The shaded area shows the
region covered by \sco\ during in all our observation (see Fig.\,7). The
position of the count rate/frequency peak (``Peak") is marked, as well as the
positions of the two count rate and colour dips (``1st Dip" and ``2nd Dip").
The ``Start" and ``End" points of the frequency excursion are also indicated.
These are points just preceding the first dip and following the second dip.
Although the Peak and Dips are consistent with other data, \sco\ does not
follow its usual track when moving to and from the peak. 

{\bf Fig.\,10.} As Fig.\,9, but with the colours instead of count rate. The
shaded area shows the region usually occupied by \sco. This area  includes the
1\,$\sigma$ uncertainties on individual points of a normal NB/FB transition on
25 Aug 1985. 

{\bf Fig.\,11.} An enlarged version of the colour-colour diagram of Fig.\,1
showing the path \sco\ took during the 13 Mar 1986 anomalous frequency
excursion. The data marked with the connected line covers exactly the same
span and is integrated in exactly the same way as the data of Figs.\,8,9 and
10. Three points are emphasized with filled symbols; the first count rate dip
(circle), the count rate peak (triangle) and the second count rate dip (square).
Note that \sco\ does {\it not} enter the flaring branch during the excursion. 

{\bf Fig\,12.} The detailed behaviour of \sco\ during the anomalous frequency
excursion with ``looping behaviour" on the 30 Aug 1983. Several times are
marked. Ip is the time of the count rate peak, Fp is the time of the frequency
peak and A and B are times when the count rate dips. The point where the
frequency drops rapidly is labeled Jump. This jump is discussed in Sect.\,3.5.
The bottom panel shows the ME Xe detector and GSPC 2--8\,keV band count
rates. In both cases the background has been subtracted. EXOSAT's pointing was 
unstable during the first 1500 sec of the observation, which started at 10:11
UT, but no corrections have been made for collimator response changes. The ME
count rates have been scaled to the full 8-detector array. The middle panel
shows the hardness ratio (HR) formed from the the ratio of the 8--15 keV and
2--8\,keV GSPC counting rates. The thick curve represents the HR as measured
on the same time intervals as used to average the power spectra. The thin line
is the hardness ratio measured, using the GSPC with an 8 sec time resolution.
The top panel shows the variation in QPO frequency. The curve follows the fits
to the power spectra. The one sigma errors are from a scans in $\chi$$^{2}$. 
The points are frequency estimates made by finding the modal frequency of
powers exceeding a preselected power level (see Fig.\,9).

{\bf Fig\,13.} The frequency-colour diagram of the anomalous frequency
excursion on 30 Aug 1983. There is a clear jump between frequencies typical of
the flaring and normal branches. A frequency-intensity diagram also shows
this.

{\bf Fig\,14.} The average speed (arbitrary units) of the motion of \sco\ along
the Z-track as a function of Z-track position (\Sz). On short (96\,sec) time
scales \sco\ moves slowest at the NB/FB vertex.

{\bf Fig\,15.} A comparison between count rate power spectra and \Sz\ power
spectra down to low frequencies.  The top two panels show the count rate power
spectra for the data taken on 11/12 March 1986 (left, mostly NB data) and 25
Aug 1985 data (right, mostly FB data). The lower two panels show the power
spectra of the \Sz\ values for the same observations, again 11/12 March 1986
on the left and 25 Aug 1985 on the right. The count rate power spectra are
normalized in the same manner as the power spectra of Fig.\,4. The expected
Poisson level has been subtracted. The \Sz\ power spectra were calculated
using the fast implementation of the Lomb-Scargle  periodogram for unevenly
spaced data (Press \etal\ 1992 and references therein). The normalization is
in terms of the data variance. The noise level, as estimated from the average
power at the highest frequencies, has been subtracted. 

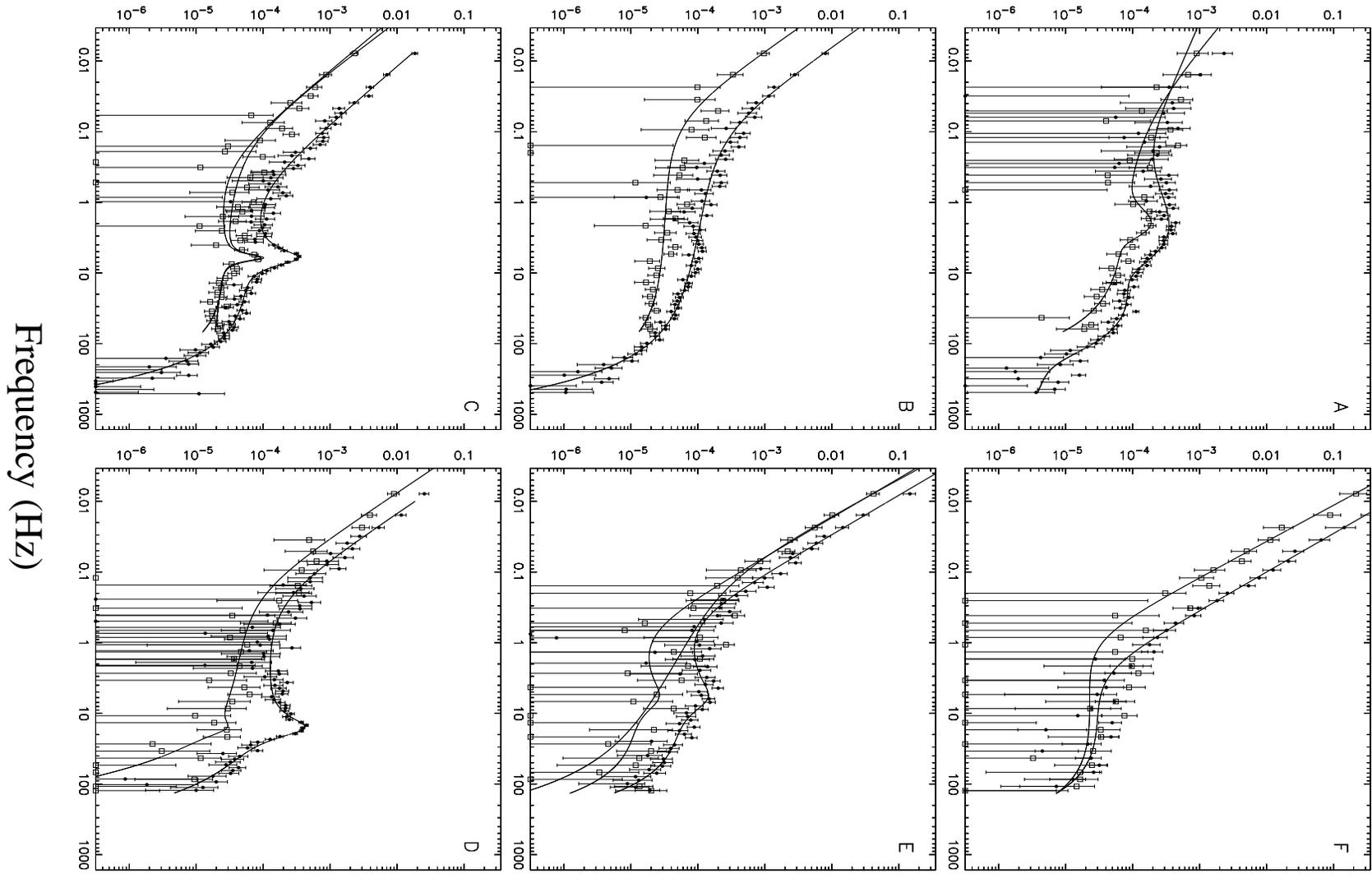
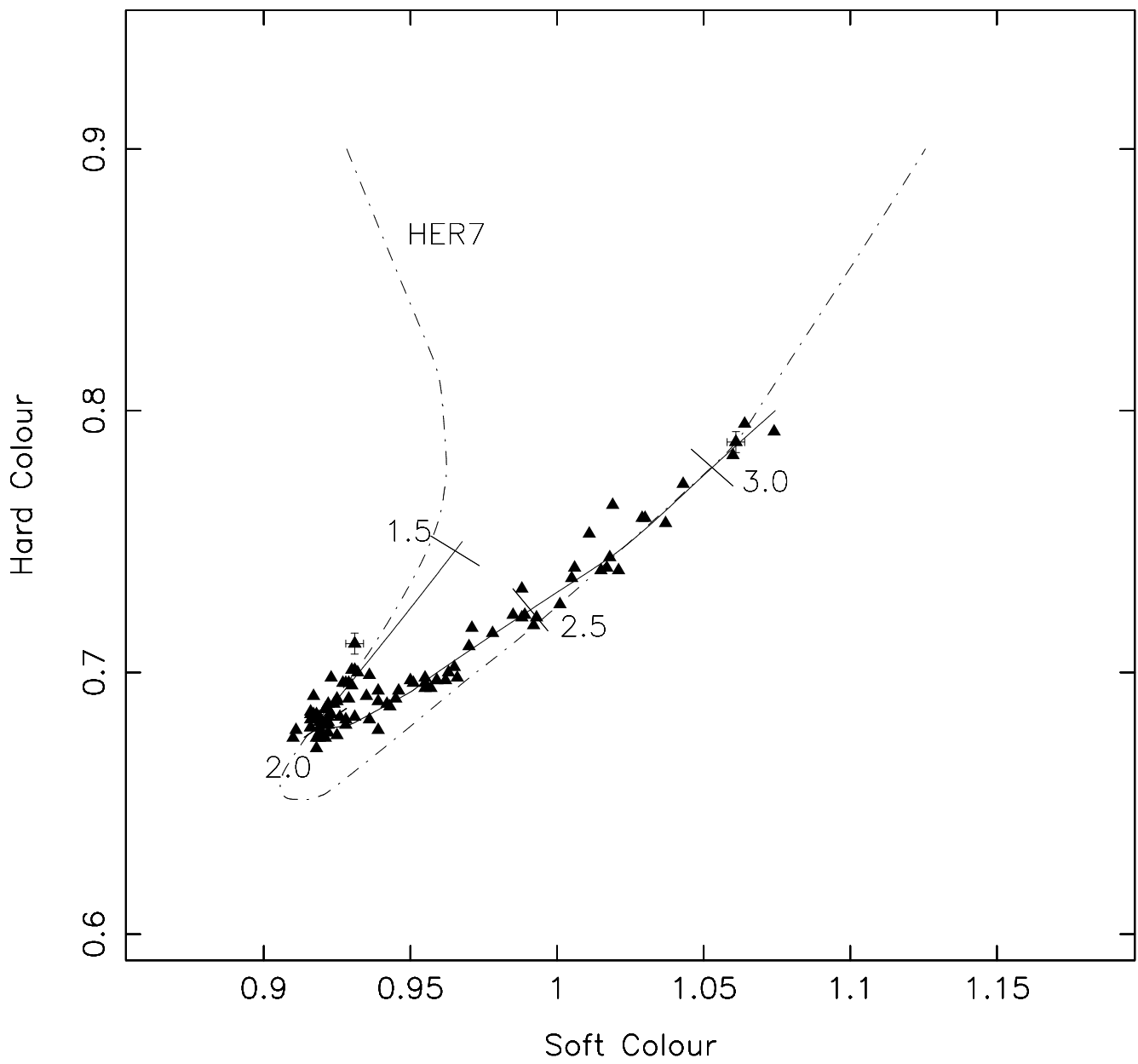
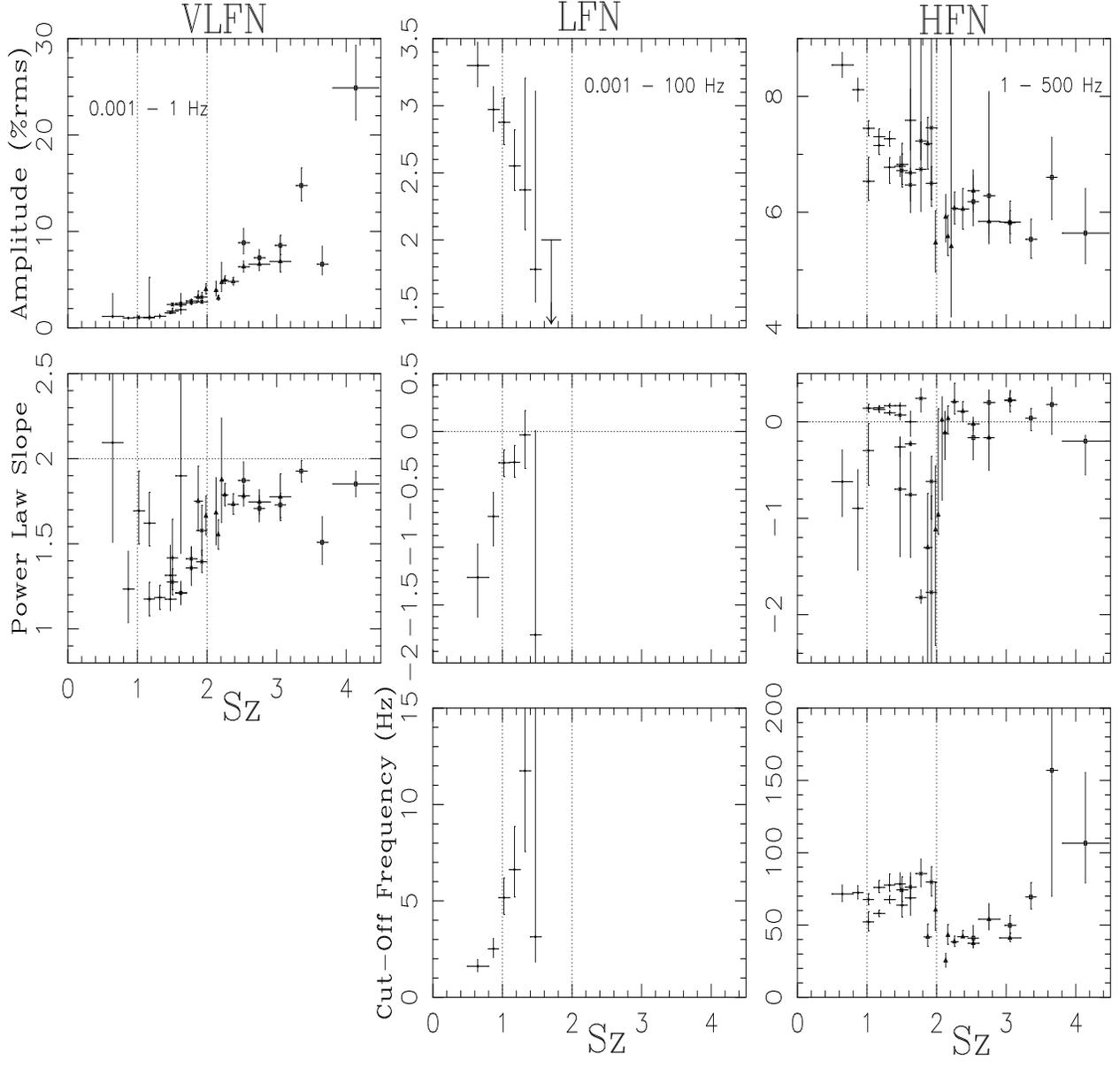
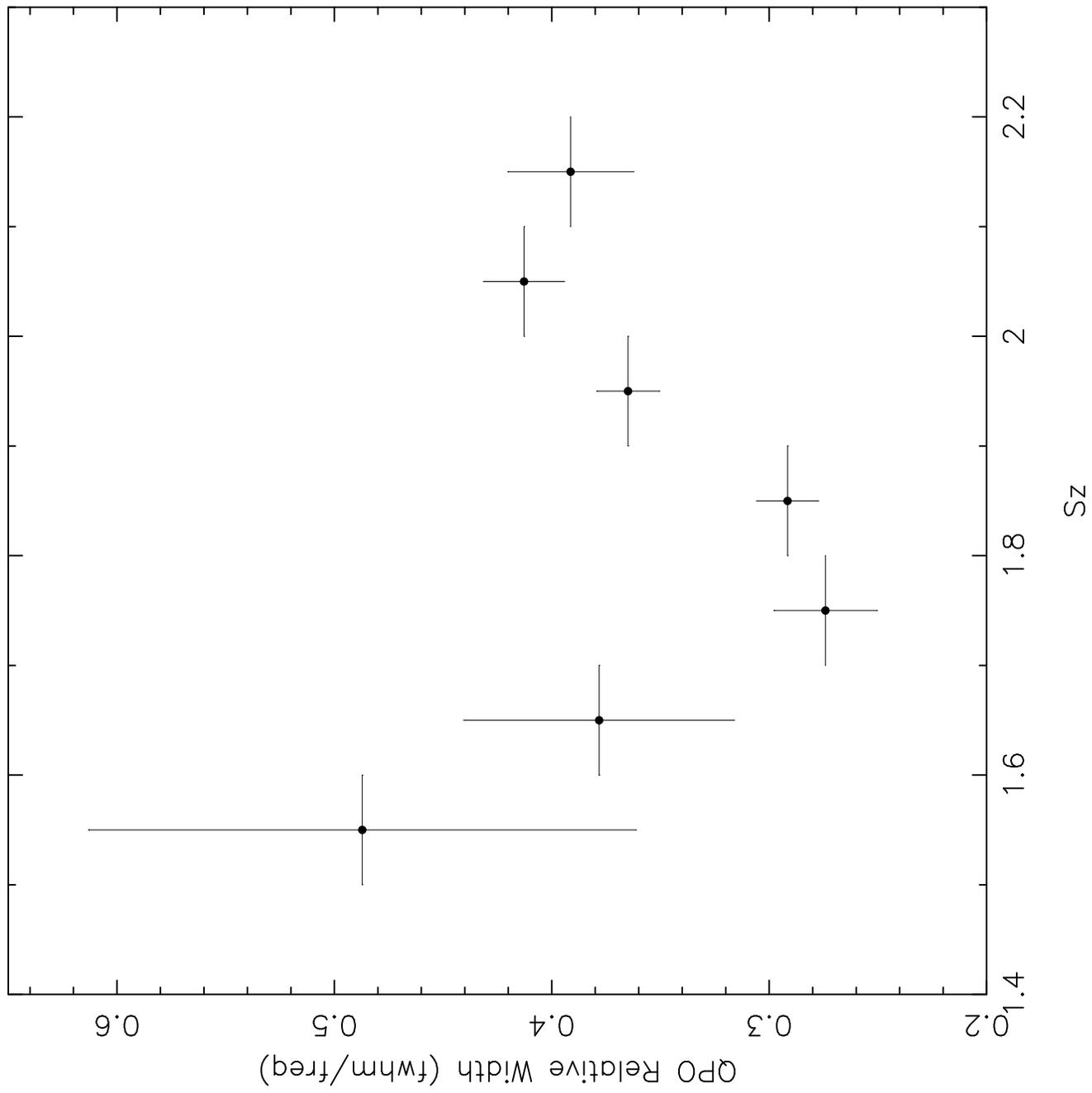
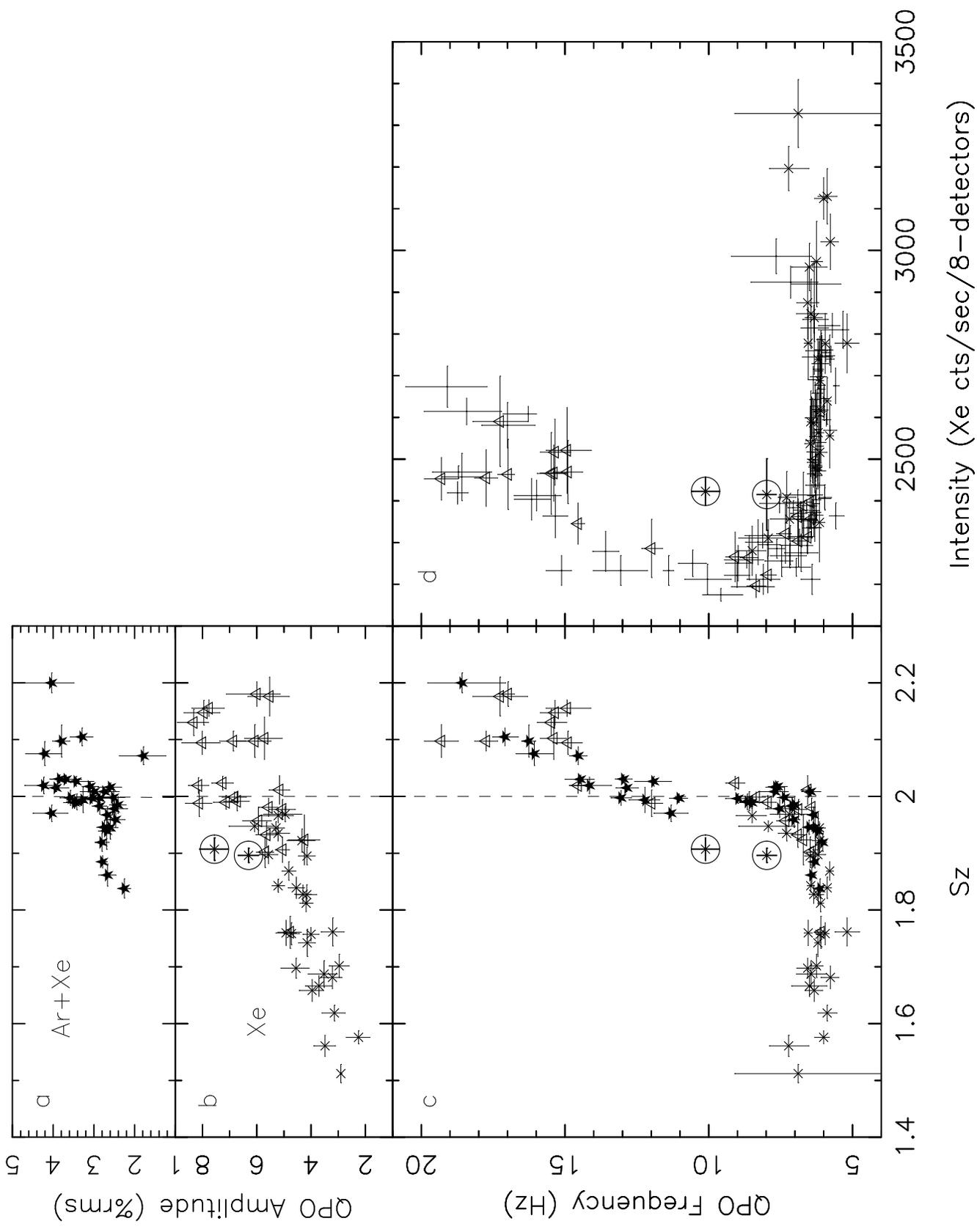
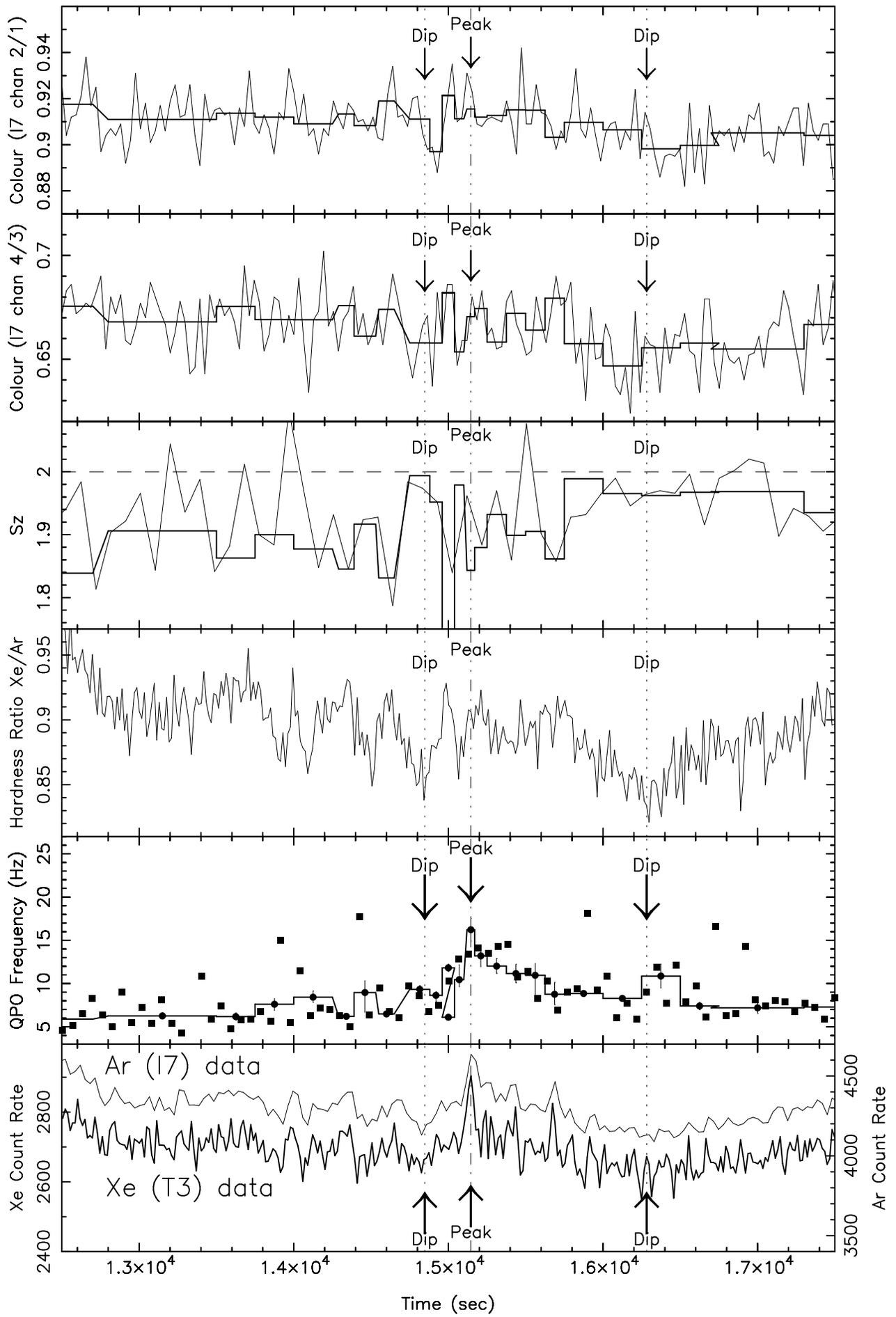
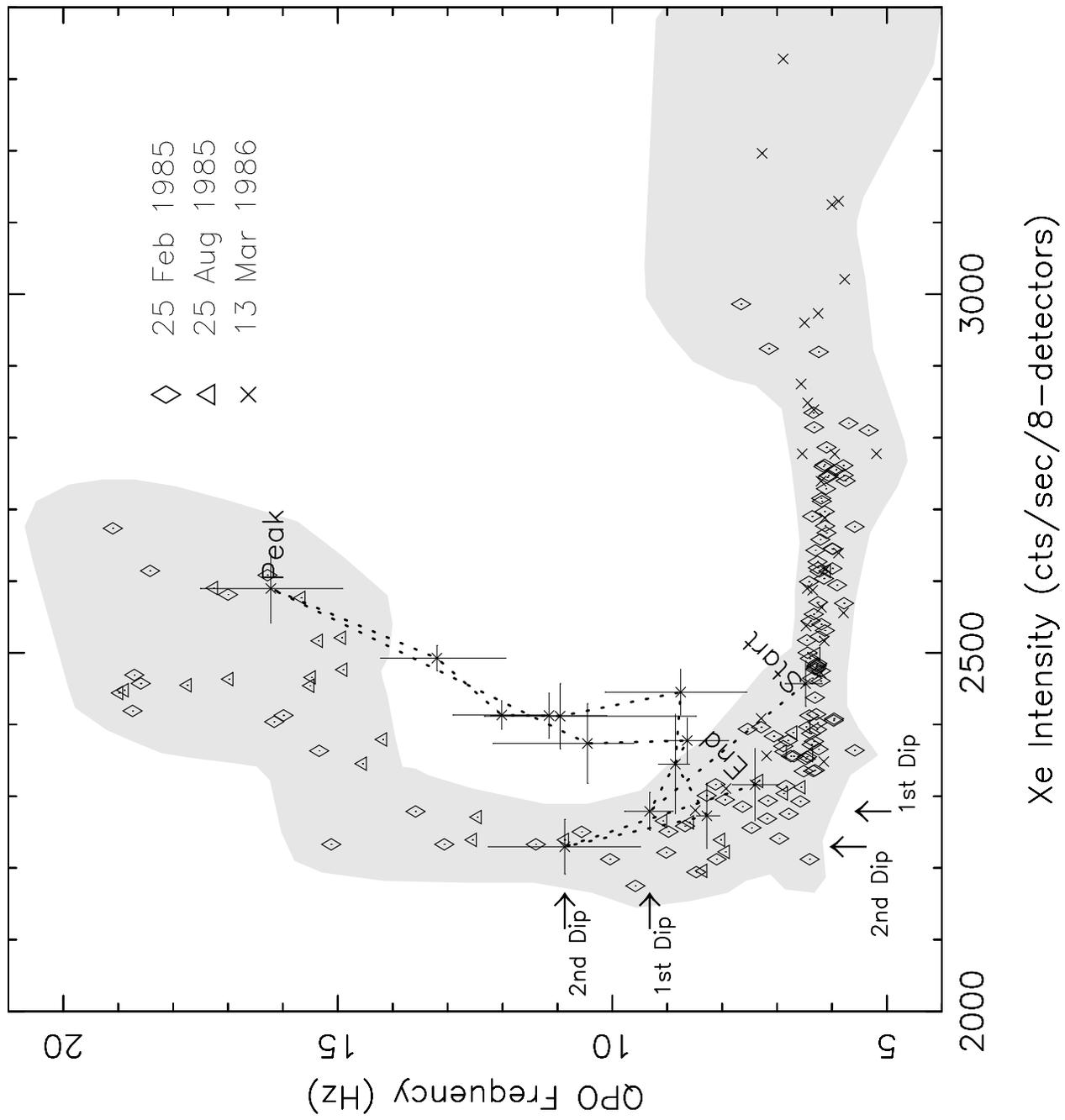
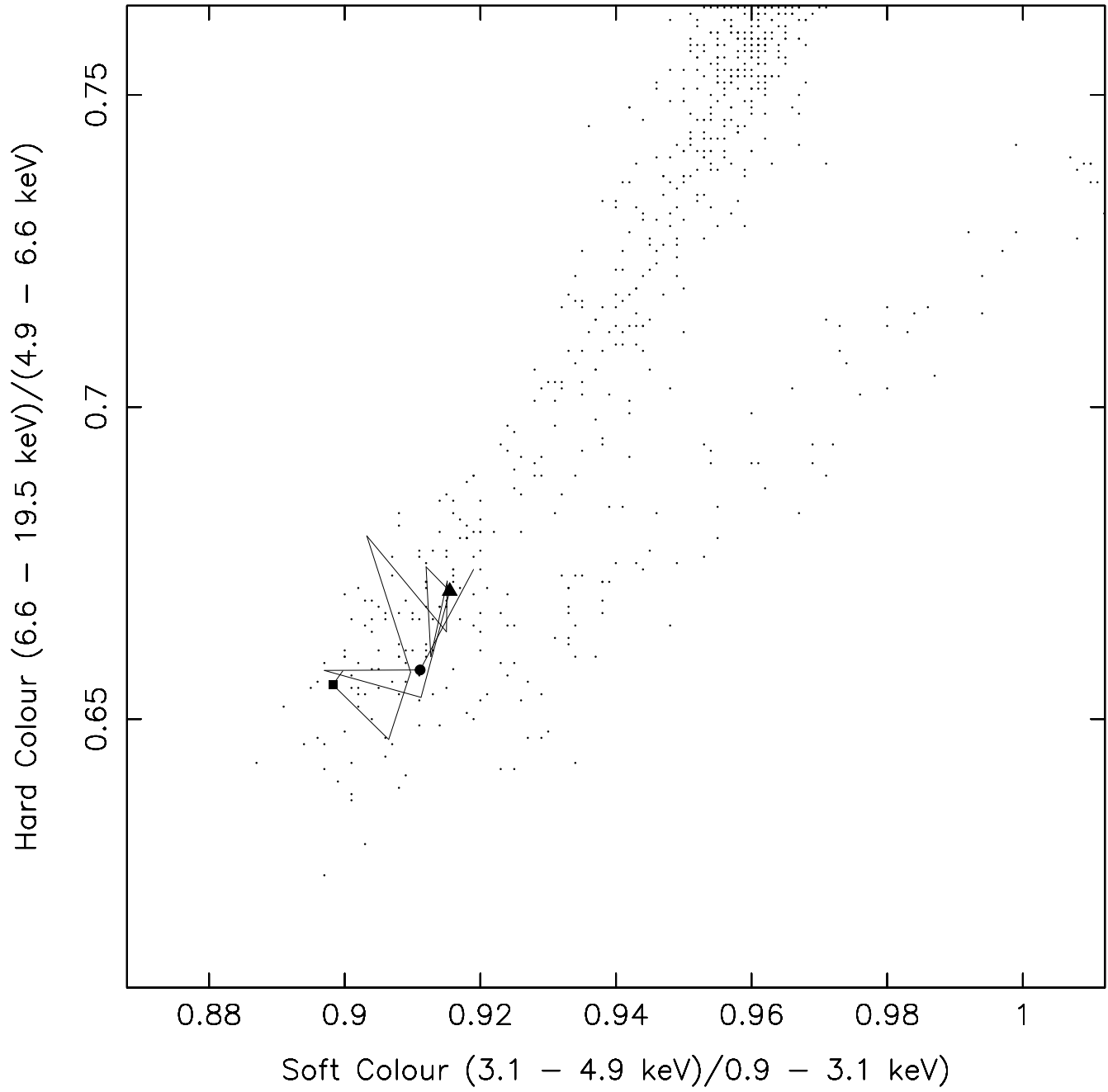
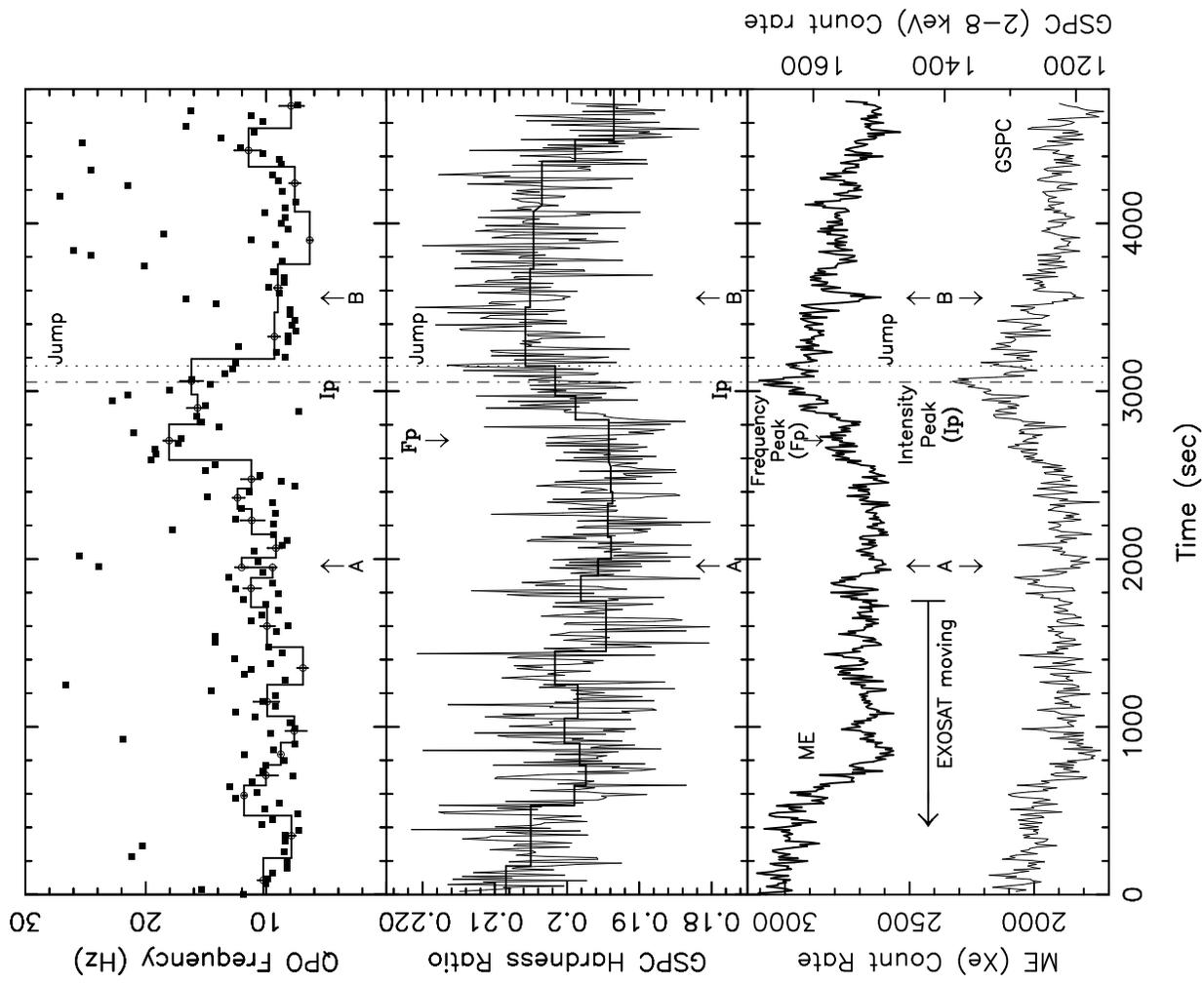
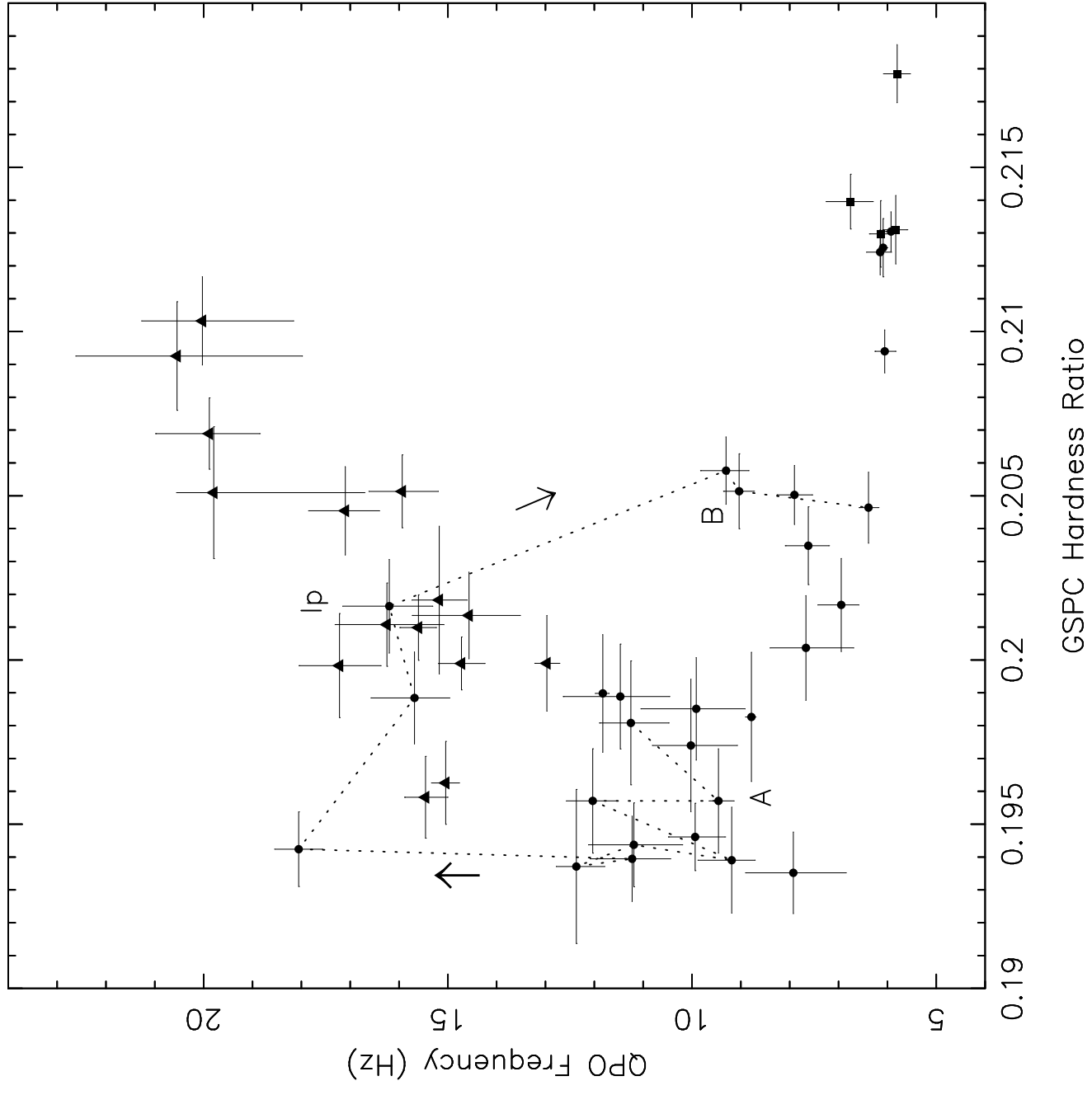
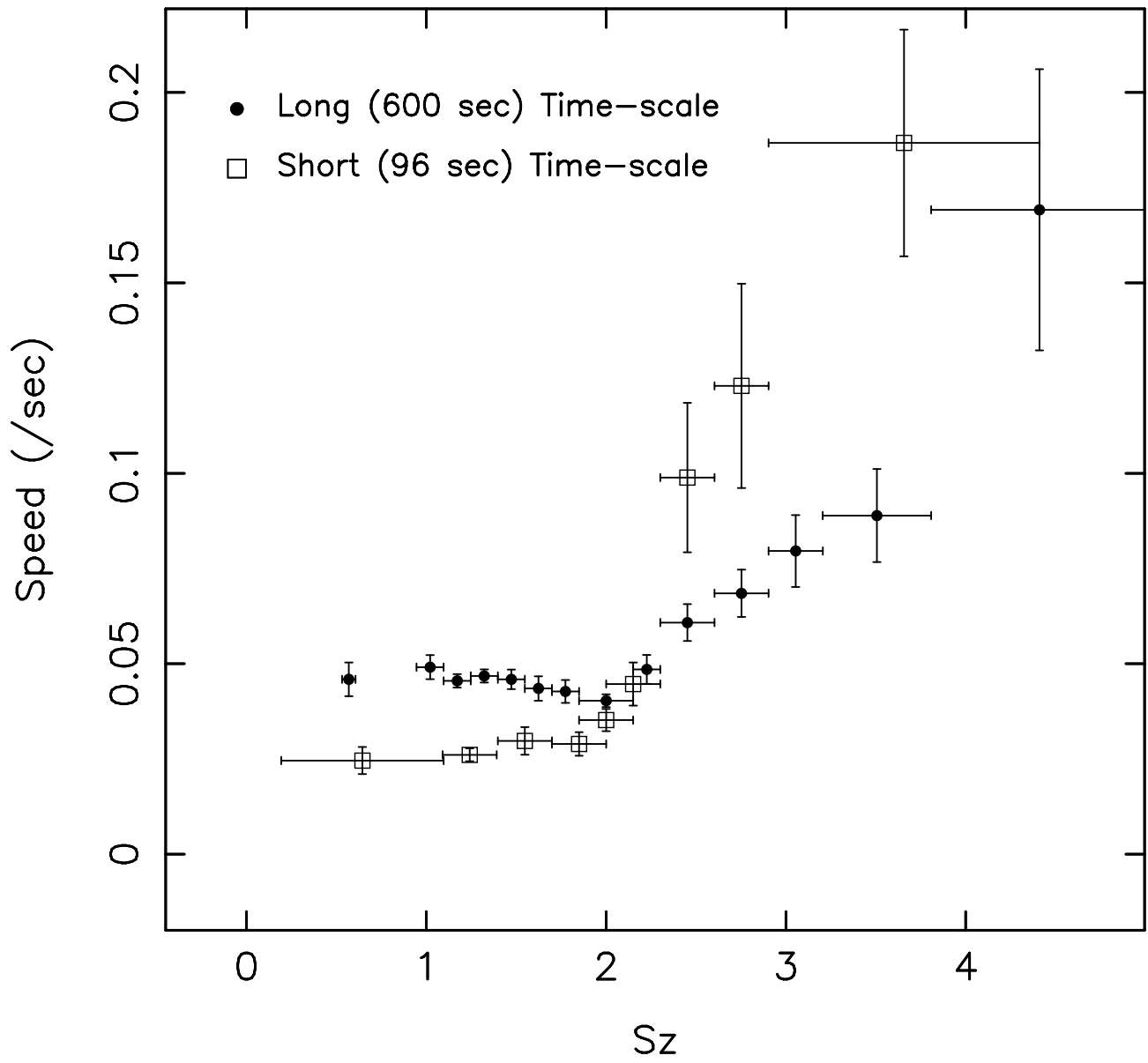
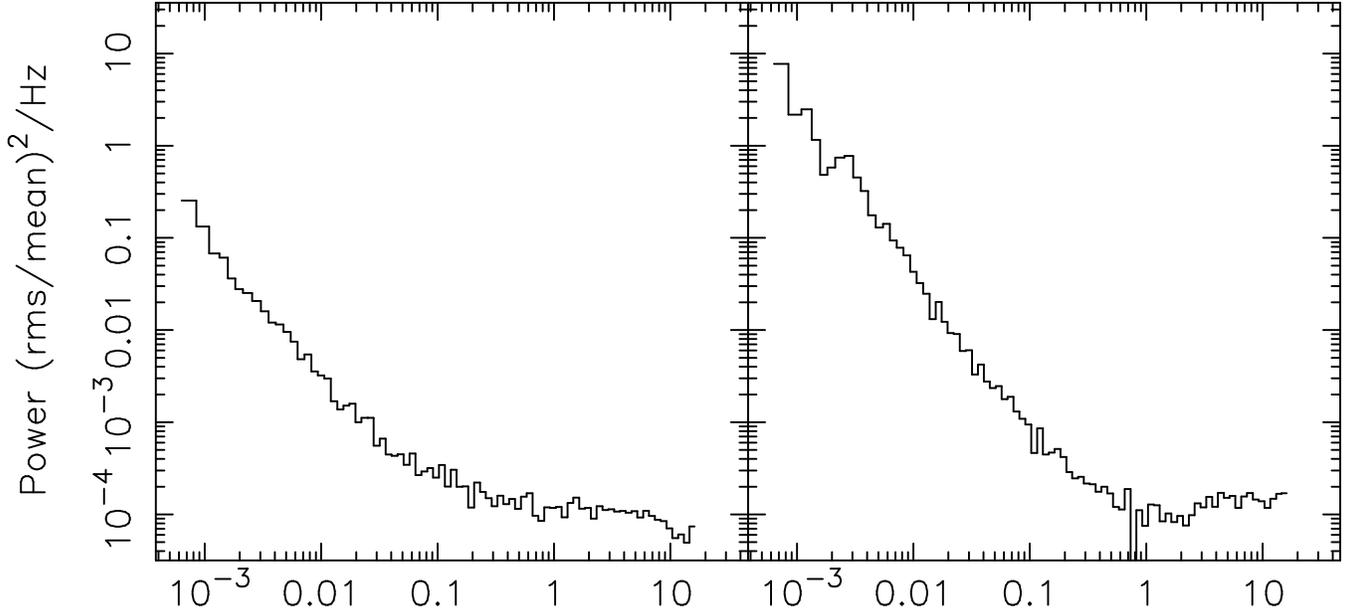
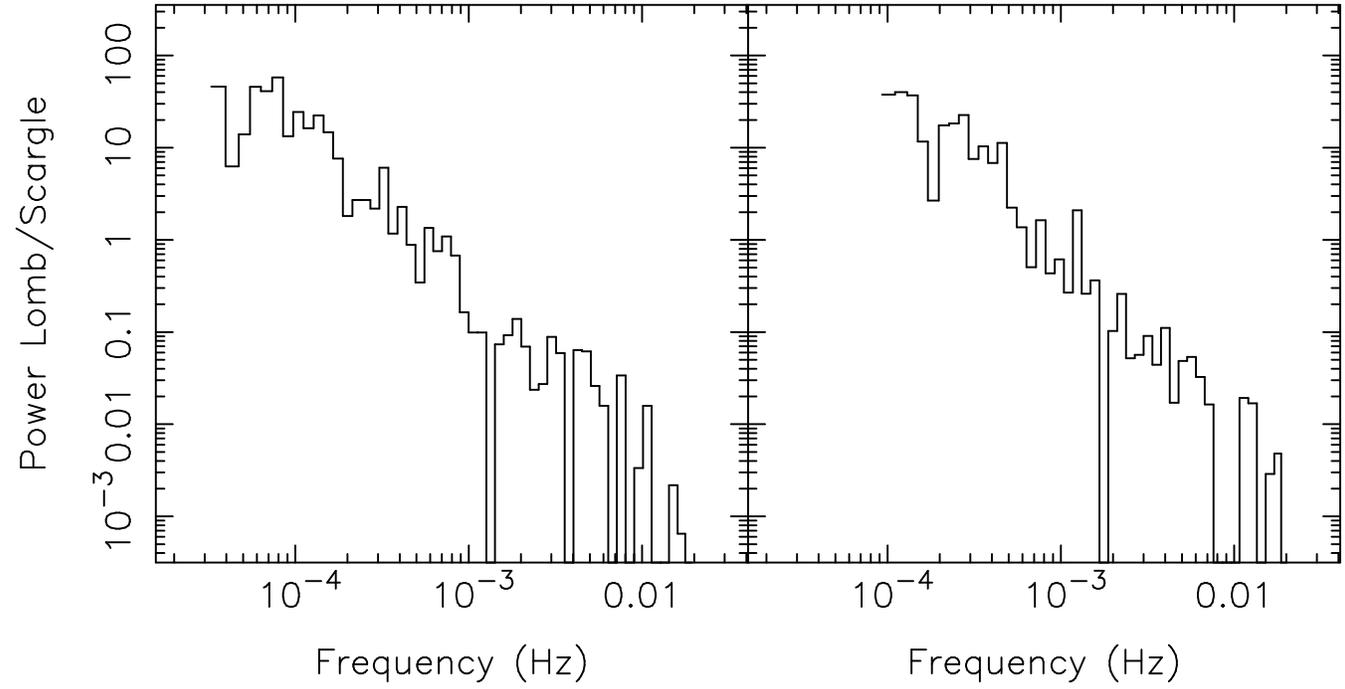
\end{document}